\documentclass[%
 reprint,
superscriptaddress,
floatfix,
 amsmath,amssymb,
 aps,
]{revtex4-2}

\usepackage{graphicx}% Include figure files
\usepackage{dcolumn}% Align table columns on decimal point
\usepackage{bm}% bold math
\usepackage{adjustbox}
\usepackage{comment}
\usepackage{ulem}

\newcommand{\updated}[1]{\textcolor{black}{#1}}
\newcommand{\figupdated}[1]{\textcolor{black}{#1}}

\usepackage[utf8]{inputenc}
\usepackage{amsmath}
\usepackage{upgreek}
\usepackage{layout}
\usepackage{mathtools}

\usepackage{xcolor}
\usepackage{graphicx}
\usepackage{amssymb}
\usepackage{siunitx}
\usepackage{float}
\usepackage{amsthm}
\usepackage{tikz-cd}
\usepackage{hyperref}
\usepackage{bbm}
\hypersetup{colorlinks=true, citecolor=blue, urlcolor=blue, linkcolor=blue, unicode=true}
\newcommand{\Z}{\mathbb{Z}}

\newcommand{\norm}[1]{||#1||}

\newcommand{\bkop}[3]{\langle #1 \vert #2 \vert #3 \rangle}
\newcommand{\ket}[1]{|#1\rangle}
\newcommand{\bra}[1]{\langle #1 |}
\newcommand{\ketbra}[2]{|#1\rangle \langle #2 |}
\newcommand{\braket}[2]{\langle #1 | #2 \rangle}

\newcommand{\mb}[1]{\mathbf{#1}}

\newcommand{\br}[1]{\left( #1 \right)}

\newcommand{\set}[1]{\{ #1 \}} % Set notation

\newcommand{\bounds}[2]{\bigg\rvert_{#1}^{#2}}
\newcommand{\boudns}[2]{\bounds} % To correct my silly spelling errors

\renewcommand{\a}{\alpha}

\newcommand{\D}{\Delta}

\newcommand{\m}{\mu}

\newcommand{\p}{\phi}
\renewcommand{\r}{\rho}
\newcommand{\s}{\sigma}

\renewcommand{\th}{\theta}

\newcommand{\W}{\Omega}

\newcommand{\bth}{{\boldsymbol\th}}
\newcommand{\bp}{\boldsymbol\p}

\newcommand{\CA}{\mathcal{A}}

\newcommand{\CF}{\mathcal{F}}

\newcommand{\CH}{\mathcal{H}}
\newcommand{\CP}{\mathcal{P}}
\newcommand{\CL}{\mathcal{L}}

\newcommand{\CN}{\mathcal{N}}

\begin{document}

% \preprint{APS/123-QED}

\title{Digital-analog quantum learning on Rydberg atom arrays}% Force line breaks with \\

\author{Jonathan Z. Lu}
\email{lujz@mit.edu}
\altaffiliation{Corresponding author. Presently at Department of Mathematics, MIT.}
\affiliation{Department of Physics, Harvard University, Cambridge, MA 02138, USA}
\affiliation{QuEra Computing Inc., 1284 Soldiers Field Road, Boston, MA 02135, USA}

\author{Lucy Jiao}%
\affiliation{Department of Physics, Harvard University, Cambridge, MA 02138, USA}

\author{Kristina Wolinski}
\altaffiliation{Presently at Department of Physics, Princeton.}
\affiliation{Department of Physics, Harvard University, Cambridge, MA 02138, USA}
%\email{kw3241@princeton.edu}

\author{Milan Kornjača}
\affiliation{QuEra Computing Inc., 1284 Soldiers Field Road, Boston, MA 02135, USA}

\author{Hong-Ye Hu}
\affiliation{Department of Physics, Harvard University, Cambridge, MA 02138, USA}

\author{Sergio Cantu}
\affiliation{QuEra Computing Inc., 1284 Soldiers Field Road, Boston, MA 02135, USA}

\author{Fangli Liu}
\email{fliu@quera.com}
\affiliation{QuEra Computing Inc., 1284 Soldiers Field Road, Boston, MA 02135, USA}

\author{Susanne F. Yelin}
\affiliation{Department of Physics, Harvard University, Cambridge, MA 02138, USA}

\author{Sheng-Tao Wang}
\email{swang@quera.com}
\affiliation{QuEra Computing Inc., 1284 Soldiers Field Road, Boston, MA 02135, USA}

\date{November 27, 2024}% It is always \today, today,

\begin{abstract}
    We propose hybrid digital-analog learning algorithms on Rydberg atom arrays, combining the potentially practical utility and near-term realizability of quantum learning with the rapidly scaling architectures of neutral atoms. Our construction requires only single-qubit operations in the digital setting and global driving according to the Rydberg Hamiltonian in the analog setting. We perform a comprehensive numerical study of our algorithm on both classical and quantum data, given respectively by handwritten digit classification and unsupervised quantum phase boundary learning. We show in the two representative problems that digital-analog learning is not only feasible in the near term, but also requires shorter circuit depths and is more robust to realistic error models as compared to digital learning schemes. Our results suggest that digital-analog learning opens a promising path towards improved variational quantum learning experiments in the near term.
\end{abstract}
\maketitle

\section{\label{sec:intro}Introduction}
Neutral atoms trapped by optical tweezers hold great potential for a scalable quantum information processing
~\cite{saffman2010quantum, Endres2016, Barredo2016, Morgado2021, Kornajca2024}. Programmable quantum simulators comprised of an array of Rydberg atoms have already scaled to several hundred qubits, enabling recent experiments that have demonstrated intriguing emergent many-body quantum phenomena including quantum phase transitions~\cite{qpt1, Scholl2021,qpt3,qpt4,qpt5,Chen2023}, many-body quantum scars~\cite{scar1,scar2,scar3}, and topological phases of matter~\cite{spinliquid1,spinliquid2,spinliquid3,spinliquid4,spinliquid5}. Rydberg atom simulators have even been shown to naturally encode hard computational problems such as the \texttt{Maximum Independent Set} problem, leading to new connections between quantum physics and \textbf{NP}-complete computational problems~~\cite{ebadi2022quantum, Kim2022,nguyen2023quantum,pichler2018computational,wurtz2022industry, Byun2022}. The rapid progress towards larger scale Rydberg quantum processors motivates us to examine their applications for near-term quantum algorithms. In the case of Rydberg system using Rubidium atoms, a simplified picture of its quantum simulator is a system of atoms trapped in a desired arrangement, individually in their ground state (denoted $\ket{0}$), that is globally driven for some finite time $t$ under the Rydberg many-body Hamiltonian~\cite{wurtz2023aquila}. This capability alone is enough to realize many exotic physical properties of Rydberg arrays discussed above, and present-day simulators may be able to simulate other complex physical models, such as those in high-energy physics~\cite{surace2020lattice,surace2021scattering}. Beyond global driving, another high-fidelity control that can be added to a Rydberg atom array system is single-qubit rotations. These gates operate in a space of hyperfine states rather than the Rydberg two-level system, which we will describe in more detail in the following sections. This ability to locally change the basis of each qubit enables a new host of applications, including quantum machine learning (QML). Recent theoretical studies have indicated that QML, in some cases, can provide practical benefits for certain problems~\cite{huang2021information,riste2017demonstration,anschuetz2023interpretable, Biamonte2017,hy1,hy2,hy3,li2022recent}. The possibility of executing quantum computations with practical utility motivates us to study explicit constructions of such learning algorithms. In this paper, we explore quantum learning algorithms from near-term Rydberg atom simulators---their construction and potential benefits over \updated{digital} variational quantum circuit models~\cite{Quek2024,Schuster2024}.

Variational quantum algorithm (VQA) is one of the most widely considered QML models. Experimentally, the VQA recipe yields algorithms generally compatible (under reasonable circuit depth) with near-term quantum devices~\cite{cerezo2021variational}. A remarkably wide class of VQA and VQA-variants have been proposed in recent years for applications that span a range from purely computational and physical problems to drug discovery~\cite{cerezo2021variational,benedetti2021hardware,du2022efficient,uvarov2021barren,self2021variational,jones2019variational,robert2021resource,liu2021variational}.

\begin{figure}
    \centering
    \includegraphics[width=\columnwidth]{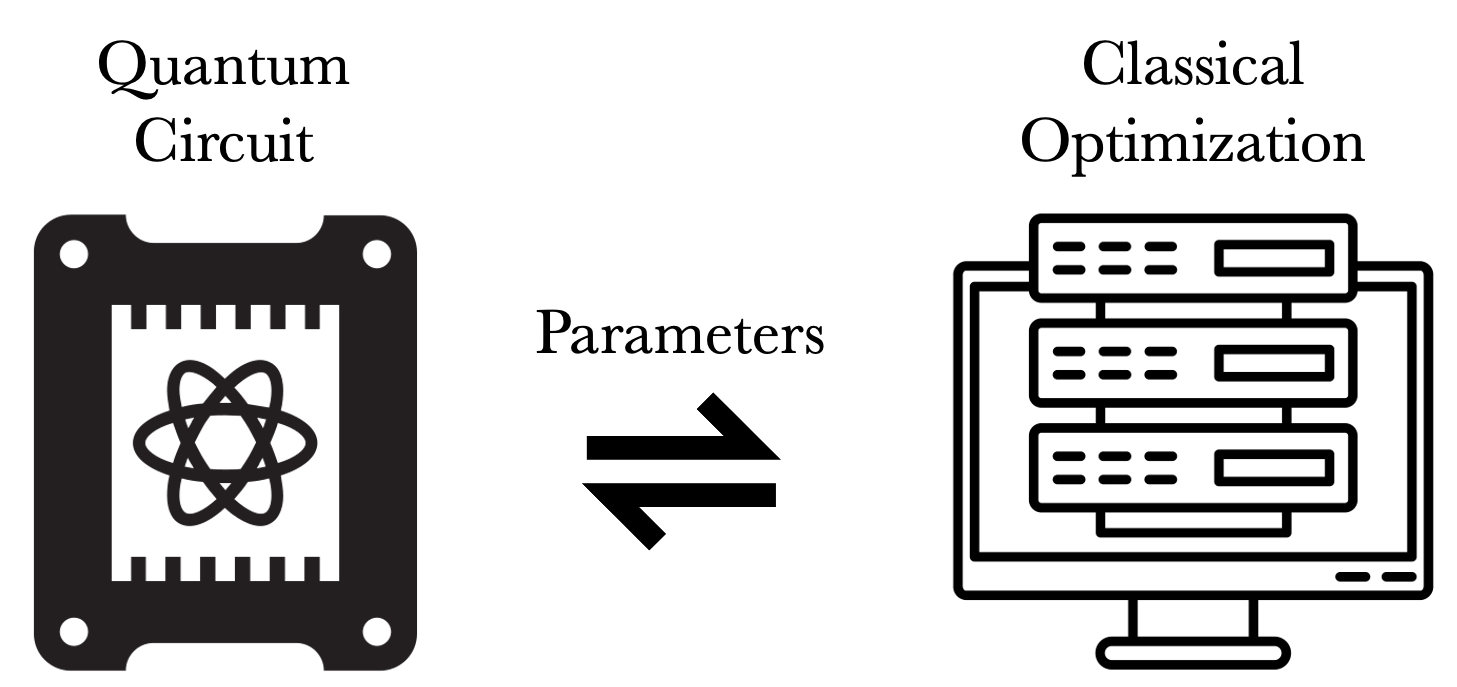}
    \caption{A VQA consists of a variational quantum circuit exchanging parameters with a classical optimization algorithm.}
    \label{fig:VQA_structure}
\end{figure}

We begin by briefly reviewing the general structure of a VQA so that we may connect it directly to our Rydberg construction. The VQA formulation follows two guiding principles: divide the tasks of machine learning between quantum and classical computers, and allocate to each resource the problems it is best at solving~\cite{cerezo2021variational}. Depicted schematically in Fig.~\ref{fig:VQA_structure}, a VQA, therefore, consists of a quantum circuit and a classical computer interacting with each other. The quantum circuit contains a set of parameters that may be varied to learn an underlying problem.

In its simplest form, a learning task involves a training and a testing component. The training portion involves optimizing the parameters of the quantum circuit with respect to a loss function $\CL$ derived from the circuit measurements. This optimization is generally done classically with, e.g., simplex search and gradient descent~\cite{spall1998overview,nakanishi2020sequential,kingma2014adam,nelder1965simplex}. Such techniques require that the classical optimizer iteratively query the quantum circuit for both its parameters and its gradient\updated{; the optimizer then updates the parameter based on the gradient value. A classical analogy to this process might be a neural network optimized by gradient descent via the same iterative query process, though a VQA does not have any classical neural networks in its construction.} The testing portion consists of querying the optimized quantum circuit with new input data and analyzing the output. 

\subsection{VQAs with Rydberg atom system}

\begin{figure*}[ht!]
    \centering
    \includegraphics[scale=0.5]{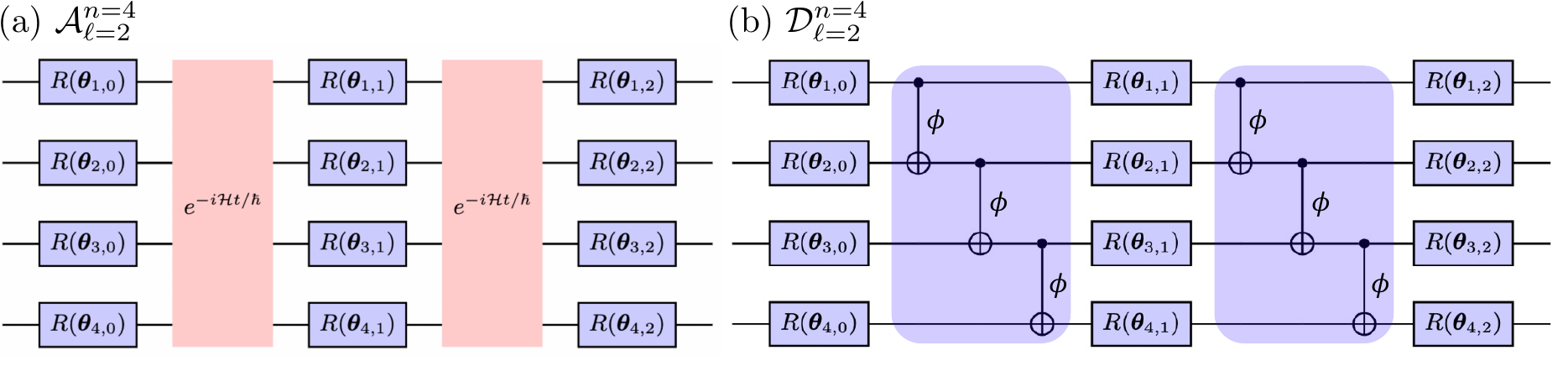}
    
    \caption{Visualization of (a) $\CA_2^4$ assuming a time-independent Rydberg Hamiltonian $\CH$ and (b) $\mathcal{D}_2^4$. Blocks shaded in blue are implemented digitally in the hyperfine ground state manifold, while blocks shaded in red are implemented by analog global driving in the ground-Rydberg two-level system.}
    \label{fig:DAQC}
\end{figure*}

One important task to be addressed on the quantum circuit side is the choice of parameterized circuit; that is, choosing an Ansatz that is most suitable for learning. A trainable variational Ansatz typically is comprised of (a) gates corresponding to a collection of optimizable parameters and (b) gates that generate entanglement. These gates need not be disjoint generally. Correspondingly, in our model of learning on globally-driven Rydberg atom arrays, we have two resources: (a) single-qubit rotation gates $R(\bth)$ with Euler angles $\bth$ and (b) global time evolution $e^{-i \CH t/\hbar}$ under the Rydberg Hamiltonian $\CH$ (see Eq.~\eqref{ryd_eq}). Gates of the former type (as well as any discrete multi-qubit gates) are referred to as \textit{digital} gates while global driving operations are referred to as \textit{analog} gates. \updated{ Analog gates are derived from the native Rydberg Hamiltonian dynamics, and while they do not correspond to an arbitrary unitary on the qubits they are applied on, we explore their effectiveness as a QML entangling layer.} A general family of Ans\"atze (parameterized over system size $n$ and circuit depth) assuming a time-independent Hamiltonian is given by a recursive definition over the number of layers $\ell$: \begin{align}
\begin{aligned}    \CA_\ell^n\left(\set{\bth_{i,j}}_{i=1, j=0}^{n, \ell} \right) & = \CA_{\ell-1}^n\left(\set{\bth_{i,j}}_{i=1, j=0}^{n, \ell-1}\right) \cdot e^{-i \CH t/\hbar} \\
    & \phantom{goodday} \cdot \bigotimes_{i=1}^{n} R_i\left(\bth_{i,\ell}\right)
\end{aligned}
\end{align}
with a base case $\CA_{0}^n\left(\set{\bth_{i,0}}_{i=1}^{n}\right) = \bigotimes_{i=1}^{n} R_i\left(\bth_{i,0}\right)$. The number of layers $\ell$ is equivalently the number of time evolution gates, such that the depth of a $\CA_\ell^n$ is $d = 2\ell + 1$. An example for $n = 4, \ell = 2$ is given in Fig.~\ref{fig:DAQC}(a). All circuits of the form $\CA_\ell^n$ are referred to as hybrid digital-analog (DA) learning circuits.

In this paper, we conduct a thorough numerical study of the performance and limits of practical quantum learning on digital-analog circuitry that can be feasibly implemented in Rydberg atom arrays. To analyze the capabilities of digital-analog learning under realistic conditions, we compare digital-analog circuits with a natural digital counterpart $\mathcal{D}_\ell^n$, shown in Fig.~\ref{fig:DAQC}(b). \updated{In $\mathcal{D}_\ell^n$}, each time evolution operator is replaced with a layer of $n-1$ adjacent generalized controlled-NOT gates \updated{between nearest neighbors. These generalized gates are endowed with a global parameter $\phi$ such that the gate is precisely CNOT---mapping basis elements $\ket{b_1, b_2} \mapsto \ket{b_1, b_1 \oplus b_2}$---when $\phi = \pi/4$, but can be more general. The extra parameter gives additional power to the digital scheme, which in turn strengthens the performance gap between digital and digital-analog schemes overall.}

%The constraint to nearest-neighbor interactions is reflects the near-term capabilities of a digital quantum machine (though in principle one could consider next-nearest-neighbors and so on) and the choice of the controlled-NOT gate as the entangling operator is conventional; any entangling two-qubit gate is equivalent to CNOT up to single-qubit gate adjustments. 

In order to conduct a comprehensive comparative numerical study, we choose two representative learning problems on which we apply digital-analog learning. The first is a supervised learning problem where the input is classical information encoded into a product state: recognition of handwritten digits from the MNIST dataset~\cite{deng2012mnist}. The second is an unsupervised learning problem where the input is quantum information: learning phase boundaries of various Hamiltonians based on the ground state as input. Both of our algorithms can be directly implemented on a Rydberg atom array that admits global driving/time evolution and single-qubit operations. We will show that in both cases, digital-analog methods have benefits over purely digital techniques in performance and in robustness to error, poising them as a promising technique to execute VQA's in the near term.

In Section~\ref{sec:rydberg}, we describe the time evolution generated by the Rydberg Hamiltonian and its relation to experimental constraints, and discuss realistic noise models for both digital and digital-analog circuitry. In Section~\ref{sec:results}, we discuss our results from learning both classical and quantum information. In Section~\ref{sec:conclusion}, we summarize our work and discuss an outlook on digital-analog learning.

\subsection{Related work, contributions, and limitations}
The idea of digital-analog circuitry has been proposed in the literature. Hybridizing single-qubit digital gates and analog evolution has been studied for quantum simulation purposes under simple models, such as the Ising Hamiltonian for the simulation of spin models~\cite{parra2020digital,garcia2023digital}. Numerical simulations of quantum Fourier transforms and phase estimation algorithms implemented with digital-analog architectures have given evidence of their comparative robustness to noise~\cite{martin2020digital,garcia2021noise}. It is even possible that digital-analog quantum computation may open an avenue to a realizable practical quantum advantage~\cite{daley2022practical}.

Our primary contributions are twofold. First, we provide a comprehensive comparative study of digital-analog architectures on representative learning problems spanning classical and quantum data. It is natural to study performance on both forms of data; such a juxtaposition has been considered experimentally as well~\cite{ren2022experimental}. Second, we co-design our constructions with Rydberg devices, requiring only single-qubit digital gates and global driving operations. We show explicitly that physically astute choices of analog pulse parameters lead to high-performance digital-analog circuits.
The current state of the Rydberg platform moreover suggests that our proposed learning techniques can be implemented within the next few years on Rydberg devices at the order of hundreds of qubits or more~\cite{wurtz2023aquila,qpt5,scar1,spinliquid1, Kornajca2024}. Time evolution on Rydberg atom arrays has been extensively studied, and high-fidelity single-qubit gates can be implemented at scale via hyperfine state manipulation~\cite{hyper1,hyper2,hyperfine, Evered2023}. Consequently, we propose these architectures for the purpose of near-term realization on Rydberg devices. At the same time, however, the robustness of our results suggests that digital-analog architectures may be beneficial more generally for QML tasks, and should be viewed as a generic technique to boost learning schemes rather than an attempt to deal with the handicap of near-term device limitations.
 
\updated{Although both schemes have hyperparameters in the entangling layer, one might argue that choosing all of the Hamiltonian parameters and the evolution time in the analog scheme are more challenging computationally than the choice of $\phi$ in the digital scheme. For example, a simple choice of $\phi$ is $\pi/4$; that is use CNOT gates.} \updated{In general, if there are many hyperparameters in a model, they may be practically challenging to optimize by brute-force computation.} However, for the case of Rydberg Hamiltonians, we are able to circumvent this problem by arguing for the optimality of specific hyperparameters guided by the physics of Rydberg atoms. These arguments, which we present in the following sections, are physically motivated and yield excellent results. Consequently, this na\"ive difficulty when guided by physics reveals itself to actually be a significant \textit{advantage} of digital-analog learning, allowing for improved learning without blindly optimizing additional parameters through the leveraging of physical reasoning. \updated{Moreover, we will show that the simple choice of $\phi = \pi/4$ is substantially worse than the optimal choice.}

% We remark that such an Ansatz is inspired by the classical neural network, which alternate parameterized linear operations and nonparameterized activations whose primary purpose is to introduce, not necessarily in a precisely controlled manner, nonlinearity. Analogously, the VQA Ansatz alternates parameterized 1-local operations and nonparameterized quench operations whose primary purpose is to introduce, not necessarily in a precisely controlled manner, entanglement. In principle the quench operations have tunable parameters, but there are physical and computational motivations for fixing them as constants, as we discuss in the following sections. This VQA Ansatz can thus be considered a quantization of deep neural networks, the simplest and most commonly studied networks. Quantization of more sophisticated variants such as convolutional neural networks~\cite{cong2019quantum} and recurrent neural networks~\cite{bravo2022quantum} have been studied but are out of the scope of this work.
% -------------

\section{\label{sec:rydberg}Theory}
We consider a model of a Rydberg system that is described by two distinct modes; the device can switch between the two modes by certain laser pulse sequences. In the digital mode, an atom in the Rydberg array can be in any one of $d$ states and is manipulated according to the standard circuit model. In this paper, we set $d = 2$ and choose two states from the Rydberg hyperfine ground state manifold to encode the digital qubit~\cite{hyper1,hyper2}. In the analog mode, each atom is described by a two-level system consisting of a ground state $\ket{g}$ and an excited state $\ket{r}$. The dynamics of the Rydberg atom array are governed by the Rydberg Hamiltonian~\cite{wurtz2023aquila}
 \begin{align}
\begin{aligned}
    \frac{\mathcal{H}(t)}{\hbar} & = \sum_{j=1}^{n} \frac{\Omega_{j}(t)}{2} (e^{i \phi_{j}(t)} \ketbra{g_j}{r_j} + e^{-i \phi_{j}(t)} \ketbra{r_j}{g_j}) \\
    & \phantom{hii} - \sum_{j=1}^{n} \Delta_{j}(t) \hat{n}_{j}+\sum_{j<k} V_{j k} \hat{n}_{j} \hat{n}_{k}
    \label{ryd_eq}
\end{aligned}
\end{align}
where $\W_j$ is the Rabi frequency of atom $j$, $\p_j$ is a phase factor, $\D_j$ is the laser detuning frequency, $\hat{n}_j = \ketbra{r_j}{r_j}$ is the Rydberg projection operator, and \begin{align}
    V_{jk} = \frac{C_6}{|\vec{r}_j - \vec{r}_k|^6} 
\end{align}   
is a van der Waals interaction potential with $C_6 = 862690 \times 2 \pi \text{ MHz } \m\text{m}^6$~\cite{wurtz2023aquila}. \updated{The detuning $\D_j(t)$, frequency $\W_j(t)$ and atomic distance $|\vec{f_i} - \vec{r_j}|$ are all parameters of experimental control.} The time dependence of the Rydberg Hamiltonian has been used for techniques such as adiabatic evolution of ground states~\cite{wild2021quantum,ebadi2022quantum}. We will discuss the role of adiabatic evolution for the preparation of ground states as quantum training data in the following section. \updated{However, to obtain the simplest operations with a tractable number of hyperparameters in our digital-analog model, we restrict to the time-independent setting.} \updated{We further simplify by enforcing site independence; that is, we} let all parameters take the same value across all atoms, i.e. $\W_i = \W_j = \W, \; \forall i, j$ and similarly for the other parameters, and set $\p_j = 0$.

As an operator, $\CH$ depends both on the geometry of the underlying Rydberg atom array and a dimensionless detuning parameter $\D/\W$. We adopt a simple geometry---a one-dimensional chain of atoms with open boundary conditions\updated{, spaced apart by a distance $a$}. The lattice parameter $a$ can be made dimensionless by introducing a characteristic length scale, the so-called blockade radius $R_b$ defined by \begin{align}
    \frac{C_6}{R_b^6} = \W
    \label{eq:blockade}
\end{align}
the length at which the interaction strength equals the Rabi frequency. The blockade radius physically corresponds to the length scale at which the interaction potential term in the Hamiltonian begins to dominate. For a ground state within this regime, among any cluster of Rydberg atoms contained in a disk of radius $R_b$, with high probability, no more than one of the atoms may be in the Rydberg state~\cite{qpt5,scar1,scar3}. 

Consequently, the Hamiltonian is fixed by a frequency parameter $\D/\W$ and a lattice parameter $R_b/a$. The evolution operation also includes a time parameter $t$. Collectively, $(\D/\W, R_b/a, t)$ define the hyperparameters over the digital-analog learning model. 

For the remainder of this paper, we will rely on context to determine whether the system is in the hyperfine or ground-Rydberg state space, and refer to both two-level systems as $\ket{0}$ and $\ket{1}$. In order to facilitate numerical simulations, we will assume that the transfer between the hyperfine and $\ket{g}-\ket{r}$ manifolds is performed ideally. This is not the case in physical realizations, as the transfer between the two manifolds is performed by single qubit pulses in the presence of non-zero Rydberg interactions. These interactions implement weakly interacting short-time Hamiltonian evolution layers before and after each single qubit rotation layer that can be absorbed in a modified entangled evolution layer. As such, they do not change the digital-analog machine learning algorithm we present, nor are they expected to significantly alter its performance. Disregarding the interacting evolution during manifold transfer allows us to treat our atoms as an effective two-level system instead of three-level, thus leading to significant computational savings. \updated{The physical implementation of qubits, however, is different in the digital rotation and analog entanglement stages of our proposal.} A brief discussion of experimental realizations of a Rydberg digital-analog system is given in Appendix~\ref{app:experiment}.

\subsection{Noise models} \label{noise_model}
To simulate realistic conditions on a Rydberg device, we adopt noise models that align with results from experimental measurements on \texttt{Aquila}, a 256-atom programmable Rydberg simulator~\cite{wurtz2023aquila}. In this work, we ignore the infidelity of single-qubit gates as they generally have much higher fidelity than multi-qubit gates~\cite{bluvstein2022quantum, Evered2023}. For \updated{feasibility of numerical simulation}, we only consider a model of coherent errors independent among every time evolution operation that captures noise in the Rabi frequencies and detunings as well as position perturbations of each atom. These are given as \begin{enumerate}
    \item noisy detuning $\widetilde{\D} \sim \D + \CN(0, 0.1 ~\textrm{MHz})$, % Note that this is not the original units we wanted, but this is what we have in code right now
    \item noisy Rabi frequency $\widetilde{\W} \sim \W \, \cdot \CN(1, 0.01)$ 
    \item and perturbed atomic coordinates according to a Gaussian process $\widetilde{r}_j \stackrel{\text{iid}}{\sim} r_j + \CN(0, 0.1 \,\m\text{m})$
\end{enumerate}
where $\CN(\m, \s)$ is a Gaussian distribution with mean $\m$ and variance $\s^2$. These noise parameters are consistent with those of Ref. \cite{wurtz2023aquila}. 
\updated{Although this noise model does not capture all possible sources of noise, it includes a large and important subset of experimental errors and provides a fair comparison of digital versus digital-analog algorithms within our scope.}

\updated{To define the digital noise model, we first establish the generalized form of the CNOT gate.} The general recipe to obtain a gate is to write it as $U = e^{-i \th H_U / \hbar}$ for a corresponding Hamiltonian $H_U$ and time $\th$, and to use a physical system described by $H_U$ to implement $U$. As a concrete example, one may achieve a controlled-$Z$ gate by a two-body Ising interaction with unit time evolution and interaction strength $\pi/4$, i.e. $CZ_{12} = e^{-i \frac{\pi}{4} (Z_1 + Z_2) + \frac{\pi}{4} Z_1 Z_2}$. % Rydberg devices in digital mode are described by such Hamiltonians, so this technique produces a realistic implementation of a digital multi-qubit gate. 
Similarly, we may write \begin{align}
    CX_{1 \to 2} = e^{- \updated{i} \frac{\pi}{4} (I_1 - Z_1) (I_2 - X_2)}
\end{align}
where $CX_{1 \to 2}$ is a controlled-NOT gate with the first qubit as the control. \updated{This equivalent definition of the CNOT gate is derived in Appendix~\ref{app:calculations}.}
\updated{More generally, define \begin{align}
    CX_{1 \to 2}(\phi) = e^{- \updated{i} \phi (I_1 - Z_1) (I_2 - X_2)}
\end{align}
to be the generalized CNOT gate, with $\phi \in [0, \pi/4]$.}
As a simple but realistic digital independent noise model, we sample $\th$ from a Gaussian distribution with mean \updated{$\m = \p$} and standard deviation $\sigma$ chosen such that the resulting gate fidelity is $99\%$.
%~\cite{wurtz2023aquila,levine2018high} as well as superconducting-qubit~\cite{xu2020high,kim2022high,moskalenko2022high} and trapped-ion systems~\cite{ballance2016high}. 
As with the digital-analog case, infidelities of the single-qubit gates are neglected. We estimated numerically the desired standard deviation of the noisy $CX$ gate to be $\sigma = 0.065$. Our analysis of noisy circuits introduces noise at every instance of the circuit evaluation, including both the training and the testing phases. We do not, so as to save simulation time, re-sample noise in estimation of the gradient and instead use the same noise on the loss function evaluation as well as its gradient in each step.

\subsection{Physical hyperparameters} \label{hyperparameters}
\updated{In Appendix~\ref{app:calculations}, we show by direct computation that an approximately optimal choice of $\phi$ is $\pi/8$ for the digital model with noise, while $\phi=\pi/4$---the controlled-NOT gate---is optimal in the absence of noise. We are, however, primarily interested in model comparisons in the presence of noise. Thus, henceforth, references to the noisy digital or generalized CNOT gate refer to the $\text{CX}_{1 \to 2}(\pi/8)$ gate, and we maintain the standard CNOT gate for the digital circuit without noise.}

We next consider the choice of hyperparameters associated with the digital-analog model.
$\W$ fixes the scale of the Hamiltonian, so we set $\W = 2\pi \times 4 \text{ MHz}$ which in turn determines $R_b$ by Eq.~(\ref{eq:blockade}). We argue that the remaining hyperparameters \updated{are approximately decoupled}, can be chosen \updated{primarily} through physical justifications rather than variational methods\updated{, and are universal}. These hyperparameters are constant from layer to layer in the circuit.

\updated{
The generic task of optimizing hyperparameters for a learning model is difficult and may be performed by brute-force grid search. However, in our case, we argue that the optimal value of each hyperparameter does not depend strongly on those of other hyperparameters. This decoupling occurs because each hyperparameter represents an underlying physical operation related to coupling strength or entanglement in a distinct fashion. Consequently, rather than conduct a single $M$-dimensional search for $M$ hyperparameters, we independently conducted $M$ one-dimensional searches in a coordinate-wise optimization. Furthermore, some of these searches can be simplified or eliminated by physical arguments, as we discuss below. Note that although this heuristic argument may be useful, it is not necessary, as it suffices to by any means obtain a set of hyperparameters that demonstrate substantial improvement over the digital model.
}

First, we discuss the choice of the evolution time $t$. For $t \ll 2\pi/\W$, the quench is so fast that it is close to the identity and therefore does not produce sufficient entanglement. Indeed, $t$ should be at least on the scale of $\sim 2\pi/\W$ to generate sufficient entanglement. If $t \gg 2\pi/\W$, the quench will cause most states (except for quantum scar states~\cite{scar1,scar3}) to thermalize, resulting in a loss of information. Therefore, an optimal $t$ exists in an intermediate zone. An intuitive choice is $t = 2\pi/\W$ itself because the quench passes through exactly one full Rabi oscillation, a physical timescale. Hence, we adopt this choice of quench timescale in this paper. Nonetheless, it is reasonable that any time on the scale of $2\pi/\W$ will produce a trainable Ansatz.

\updated{Next, we consider the Hamiltonian parameters $\D/\W$ and $R_b/a$. While we cannot deduce from physical arguments the exact optima, we are able to bound the range containing the optima, and then numerically determine the exact values.}
A practically realizable regime for $\D/\W$ is $[0, 4]$~\cite{qpt4}. We find similar performance for any $\D/\W$ in this region, and we choose $\D/\W = 0.8$. Likewise, a reasonable regime for is between $0$ and $1$: if $R_b \leq a$, the Rydberg atoms will be interacting so strongly that it will be impractical to transfer atoms between the hyperfine ground state manifold and Rydberg manifold. Moreover, the strong interactions in the $R_b/a > 1$ regime reduce the effective Hilbert space dimension, limiting the space the Ansatz can explore. Within this choice, however, there is a physically optimal region by the following plausibility argument. As $R_b/a \to 1$, the atoms move closer together and, therefore, interact more strongly, which should produce more entanglement as a resource for learning. However, at the same time, moving atoms closer implies that the relative effect of positional fluctuations \updated{in our noise model} becomes more prominent. 

To approximately optimize this tradeoff, we estimate the gate fidelity $\CF$ of the quench $e^{- i \CH t/\hbar}$ (for one layer), defined by \begin{align}
    \CF(U) = \int d\widetilde{U} \int d\psi \bra{\psi} U^\dag \widetilde{U} \ket{\psi}
\end{align}
where $d\widetilde{U}$ denotes the measure of the noise model and $d\psi$ denotes the Haar measure. \updated{The gate fidelity measures the average robustness of our entangling gate over all states. Our goal is to set $R_b/a$ as close as possible to $1$ while maintaining a high gate fidelity.
We expect that the gate fidelity essentially characterizes an operator's error robustness under a given error model for quantum learning tasks. This is because $\CF$ measures the fidelity between the noisy and noiseless output, averaged over random states and the error distribution. In a QML training task, as the model iteratively explores the space of operators, we expect the noisy gate to encounter many typical quantum states. Thus, the gate's performance based on the average state determines its typical performance during training. 
}

\begin{figure*}[ht!]
    \centering
    \includegraphics[width=0.9\textwidth]{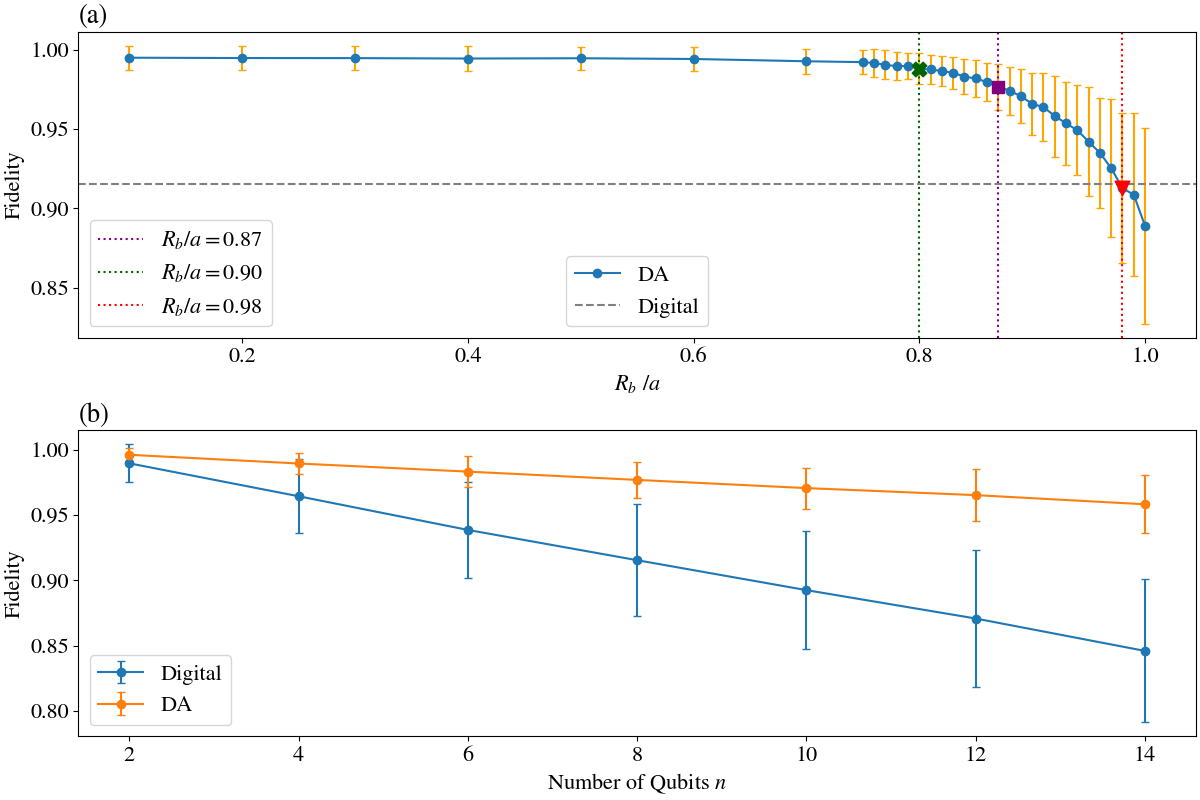}
    \caption{\figupdated{(a) Estimated fidelity of the time evolution gate on $n=8$ qubits across varying $R_b/a$. Errors bars denote the standard deviation across instances of the noisy gates, which increase as the fidelity decreases. Three points of note are $R_b/a = 0.80$ (cross), $R_b/a = 0.87$ (square), and $R_b/a = 0.98$ (triangle). The fidelity for an 8-qubit digital gate is also shown as the dotted horizontal line. Notice that the time evolution gate where $R_b/a = 0.98$ has almost the same fidelity as the digital gate.
    (b) Estimated gate fidelity, fixed at $R_b/a = 0.87$ for digital-analog circuits, compared across system sizes \updated{with that of digital circuits}.}
    }
    \label{fig:rba-fildelity}
\end{figure*}

We numerically estimate the gate fidelity by discretizing the integral over $d\widetilde{U}$ over 500 samples of the noise model and show the result in Fig.~\ref{fig:rba-fildelity}. The technique is identical for digital-analog and digital models: in each case we sample a gate from their respective noise model discussed previously and calculate the gate fidelity. Figure~\ref{fig:rba-fildelity}(a) demonstrates that the $R_b/a \to 1$ limit shows a dramatic drop in $\CF$, for a fixed $n = 8$. In particular, around $R_b/a=0.98$, the fidelity is similar to that of a digital layer. The fidelity is close to $1$ around $R_b/a = 0.8$, but we may expect---and confirm in the next section---that such a $R_b/a$ results in interactions too weak for optimal machine learning performance without noise. A reasonable trade-off is $R_b/a = 0.87$, whose approximate optimality we explore further in the next section. At this tradeoff value, the gate fidelity of the digital-analog layer is substantially larger than that of the digital layer. 

\updated{
We show in the following section that a larger gate fidelity does indeed give rise to practical performance improvements. A priori, we may already expect an advantage in the digital-analog fidelity scaling over $n$ as compared to the digital. In general, the digital gate fidelity drops quickly because we require $n - 1$ \updated{generalized} CNOT gates per layer. Since each gate is subject to an independent source of noise, the total noise should multiply, thus decreasing the error exponentially as $\sim 0.99^{n-1}$. 
% Such a scaling is evident in Fig.~\ref{fig:rba-fildelity}. 
By contrast, the analog gate, being a single global gate with a single source of error, does not have the same obvious exponential downfall and thus opens up the possibility of improved error robustness. Indeed, Fig.~\ref{fig:rba-fildelity}(b) demonstrates that this advantage exists for a range of system sizes $n$. Although we cannot determine the exact scaling of the analog gate from numerical analysis, we can see that even if it is exponential, it has a noticeably larger base than that of the digital scheme.
}

Figure~\ref{fig:rba-fildelity} also shows the standard deviation over the fidelities of the sampled set. With lower fidelities, the variance of the output between each instance of noise also increases.

\updated{
Being representative of a physical operation, each of the hyperparameters discussed above should remain optimal between various choices of the underlying computational task. We observed this independence between the two tasks we study in this work.
}

\updated{
A final hyperparameter is the number of layers $\ell$. However, $\ell$ lacks a physical interpretation and will generically be task-dependent. Specifically, $\ell$ encodes an expressivity-trainability tradeoff. If $\ell$ is too small, the model will have an insufficient number of parameters and thus lack the expressivity to perform the learning task. However, if $\ell$ is too large, then the number of errors may accumulate and result in an insufficient amount of signal to train the circuit. A large $\ell$ also suffers from higher-order effects such as potential for overfitting and an intractably large parameter dimension. In general, the optimal $\ell$ may be found by linear or binary search for a given task. We perform the former in the following section.
}

% Our results in Fig.~[\ref{fig:binary_classification}] will later confirm that $R_b/a=0.87$ balances this trade off and we will use this as the hyperparameter for our model for the rest of our results. For further evidence, we analyse over accuracies from all digit comparisons in Fig.~[\ref{fig:all-digit-compare}] in the appendix. 

% ----------
Our code used for this work~\footnote{The numerical simulations we develop in the following section were built on \texttt{Bloqade.jl} and \texttt{Yao.jl}~\cite{bloqade2023quera,YaoFramework2019} with automatic differentiation techniques that can be replaced experimentally by parameter shift rules~\cite{wierichs2022general}.} is freely available for usage~\cite{Lu_Digital-analog_quantum_learning}.

\section{\label{sec:results}Results}
We explore the effectiveness of Rydberg digital-analog learning with two different problems. In the first, we train a quantum binary classifier on  the MNIST dataset of handwritten digits~\cite{deng2012mnist}. In the second problem, we use techniques of anomaly detection to identify the boundaries of quantum phase diagrams. The input of the former problem is classical---but encoded into a product state---while the input of the latter problem is the ground state of a quantum Hamiltonian.

Note that in studying these problems we aim to provide evidence for the generic benefits of digital-analog circuitry in VQA's. Hence, we choose the given problems not necessarily for their usefulness in themselves (though phase learning can be of useful physical interest, which we discuss below), but for their representation of practical classical and quantum learning problems, respectively. Indeed, the construction of a problem for which a VQA is indisputably superior to the best classical alternative remains an open problem and is out of the scope of our present work. We remark that while our work also does not attempt to explicitly mitigate the barren plateau problem of VQA's, our loss functions consist only of local observables so that non-barren plateaus remain a possibility~\cite{cerezo2021cost}.

\subsection{\label{sec:subsec:MNIST}Digit classification}
\begin{figure*}[ht!]
    \centering
    \includegraphics[width=0.7\textwidth]{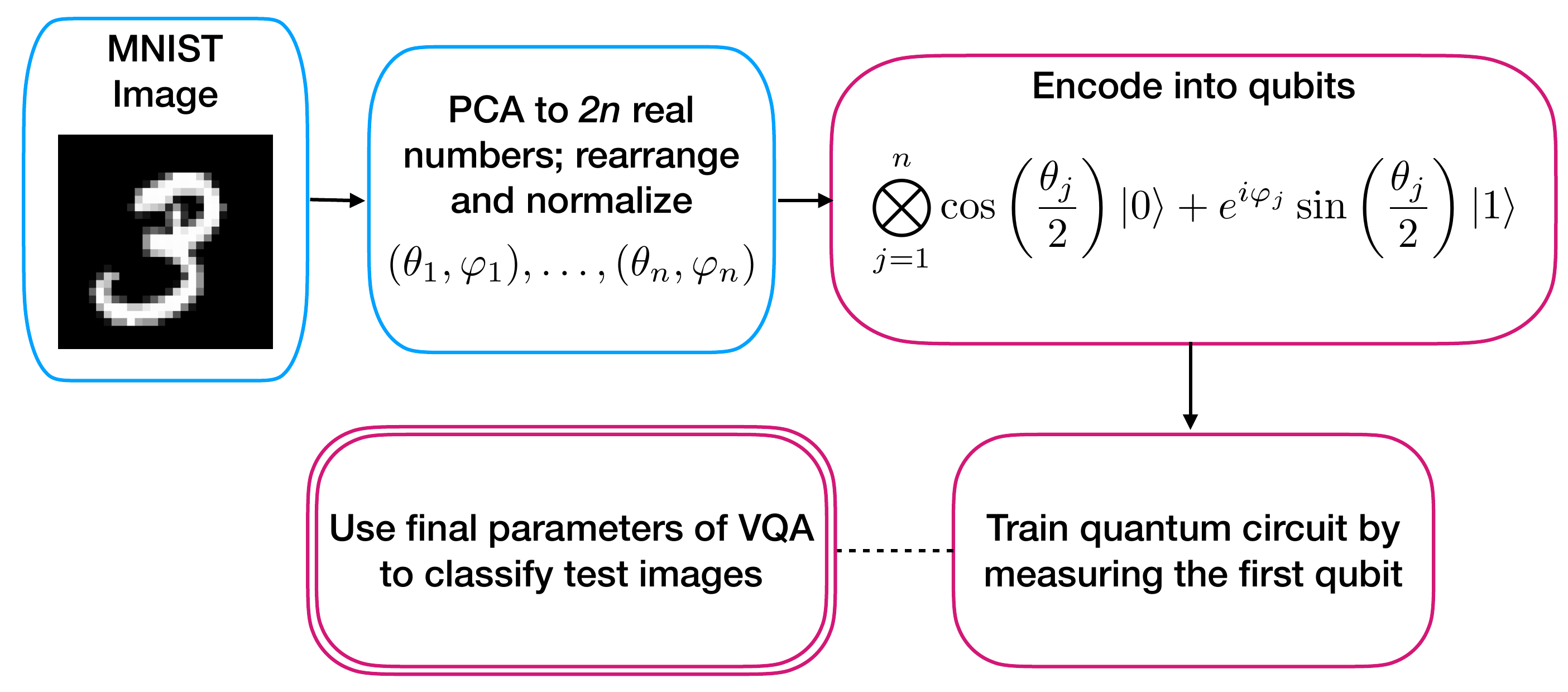}
    \caption{VQA workflow of the quantum digit classification protocol. Blue boxes are classical; pink boxes are quantum.}
    \label{fig:mnist_workflow}
\end{figure*}

The MNIST dataset contains monochromatic images of handwritten single-digit numbers and is widely used as a machine learning benchmarking dataset.
To downsample images of the MNIST dataset into a suitable form for quantum circuitry, we transform the data via principle component analysis (PCA) into a set of $2n$ numbers, where $n$ is the number of qubits. Our workflow is outlined in Fig.~\ref{fig:mnist_workflow}.
% Generally, the usage of classical pre- or post-processing on a VQA weakens the evidence of a quantum advantage since it is unclear how much of the hard work was done by the classical component. However, our goal is not to explore the advantage of VQA's; rather, we study the advantage of digital-analog VQA's over digital VQA's on a sufficiently generic classical dataset. 
The PCA output values are normalized so that they represent rotation angles, which are then encoded into product states of the form
\begin{align}
    \ket{\psi_{\text{in}}} = \bigotimes_{j=1}^{n} \; \cos\br{\frac{\th_j}{2}} \ket{0} + e^{i \varphi_j} \sin\br{\frac{\th_j}{2}} \ket{1}  .
\end{align}
Note that the principle components are ordered by importance, that is, the variance of the dataset projected along the principle component's corresponding eigenvector. We assign the first $n$ components to the $\theta$ angles and the last $n$ to the $\varphi$ angles.
We then train the variational circuit by gradient descent methods on the measurement output of the first qubit, using a local cross-entropy loss function \begin{align}
    \CL(\mb{y}, \mb{q}) = -\frac{1}{m}\sum_{i=1}^m y_i \log q(y_i) ,
\end{align}
where $y_i \in \set{0, 1}$ labels the $i$th digit in the training batch of size $m$ and $q(y) \in [0, 1]$ is the probability that the first qubit in the circuit output is 1, given $y$ as input. To estimate $\CL$, we measure the first qubit repeatedly for each fixed $y_i$, thereby obtaining an empirical estimate of $q(y_i)$. To output a classification among the two possible digits $a$ and $b$, we say that the circuit classifies the digit as $b$ if $q(y) \geq 0.5$ and $a$ otherwise.

\begin{figure*}[ht!]
    \centering
\includegraphics[width=1\textwidth]{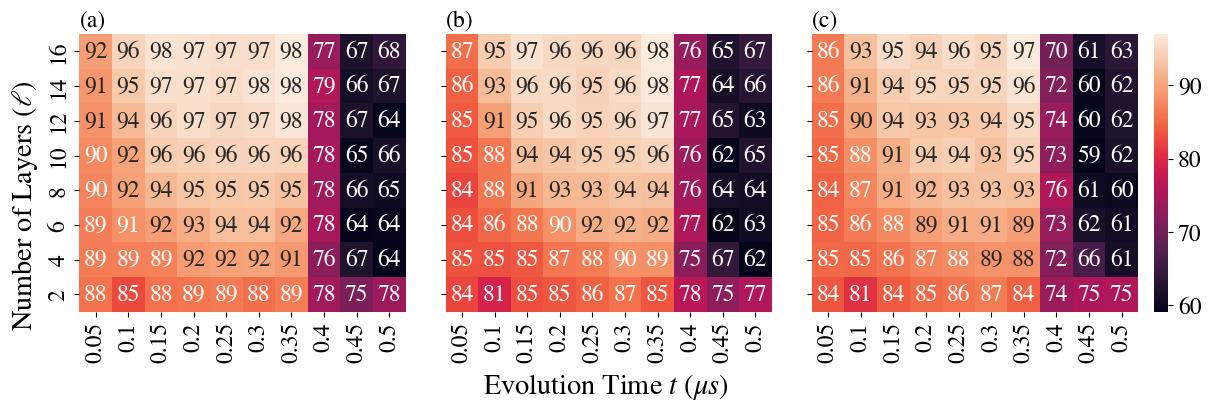}

    \caption{Comparison of accuracies over various digital-analog circuit layer depths $\ell$ and quench times $t$ in the classification of 3 versus 8, fixing $R_b/a = 0.87$, $\D/\W = 0.8$, and $n=8$ qubits. (a): noiseless model training accuracy, (b): noiseless testing accuracy, (c) noisy model testing accuracy.}
    \label{fig:binary_classification}
\end{figure*}

We begin by verifying our physical arguments in the previous section about the optimality of the quench time $t$. As an illustrative example, consider a classification between 3 and 8, intuitively a particularly difficult instance because one number is---in writing---approximately contained in the other. Figure \ref{fig:binary_classification} depicts the landscape of classification accuracies over $t$ as well as the layer parameter $\ell$, fixing $R_b/a$ and $\D/\W$. We observe that indeed near $t = 2\pi/\W = 0.25 \, \m\text{s}$, the accuracy is optimal for all $\ell$ in the absence of noise both on the training dataset shown in Fig.~\ref{fig:binary_classification}(a) and the test dataset shown in Fig.~\ref{fig:binary_classification}(b). This optimality is nearly preserved up to 1\% differences in the accuracy on the test dataset in the presence of noise, as shown in Fig.~\ref{fig:binary_classification}(c). Up to the numbers of layers we considered, increasing $\ell$ weakly improved accuracies, but at some point the introduction of additional parameters will serve only to slow training and overfit data rather than improve performance. The entire region of $t = 0.15 \text{-} 0.35 \, \m$s provides roughly equivalent performance. We proceed by fixing $t = 2\pi/\W$ and $\ell = 12$ for digital-analog analysis. This optimality is not unique to 3 and 8; for example, we show an additional comparison between 1 and 9 in Fig.~\ref{fig:087_1v9} in the Appendix.

 \begin{figure*}[ht!]
        \centering
        \includegraphics[width=1\textwidth]{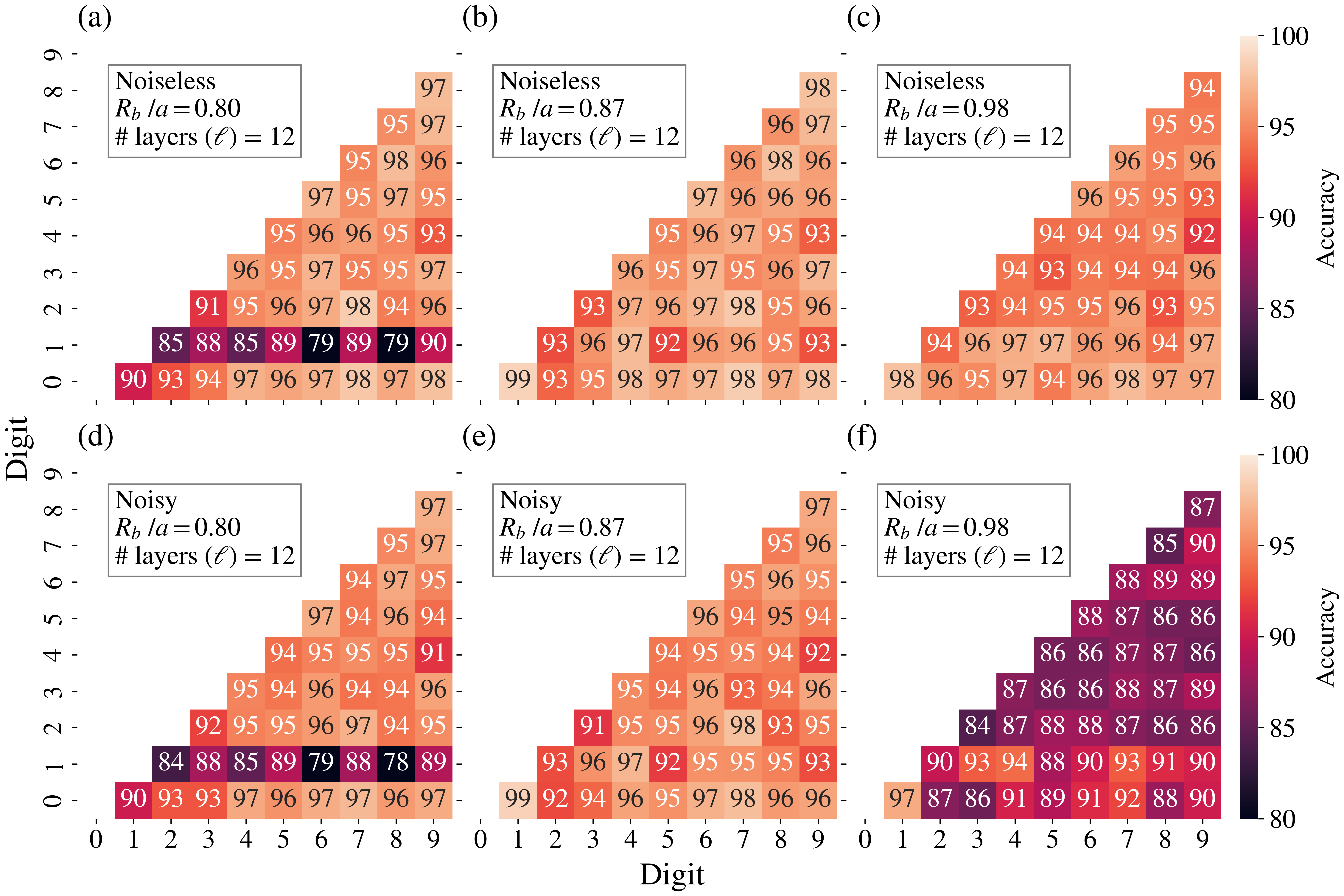}
        \caption{Accuracies for every pair of handwritten digits classified in the digital-analog model for various $R_b/a$ and fixed $t = 2\pi/\W$, $\D/\W = 0.8$, and $n = 8$. (a)-(c) neglect noise, while (d)-(f) introduce the noise model.}
        \label{fig:all-digit-compare}
\end{figure*}

Next, we verify the approximate optimality of $R_b/a = 0.87$. Figure~\ref{fig:all-digit-compare} shows choices of $R_b/a$ in the three regimes discussed in the previous section, $0.8, 0.87$, and $0.98$, for both noiseless (a-c) and noisy (d-f) instances. While $R_b/a = 0.8$ results in little ($\sim 1 \%$) loss of accuracy due to noise, the performance is quite low even without noise. On the other side of the spectrum, $R_b/a = 0.98$ fares very well in the absence of noise, but suffers heavy losses in accuracy when the noise model is introduced. The choice of $R_b/a = 0.87$ provides the predicted tradeoff between these effects, with high accuracy that is not lost in the presence of noise. While not shown here, there is a neighborhood of $R_b/a = 0.87$ that provides similar accuracies and hence small deviations around $R_b/a = 0.87$ produce similar results. The importance of this optimality is, rather, that it exists in an intermediate area between $R_b/a = 0.8$ and $R_b/a = 0.98$, but in general the choice of $R_b/a = 0.87$ suffices across different problems, such as the phase boundary problem discussed later.

\begin{figure*}[ht!]
    \centering
    \includegraphics[width=0.75\textwidth]{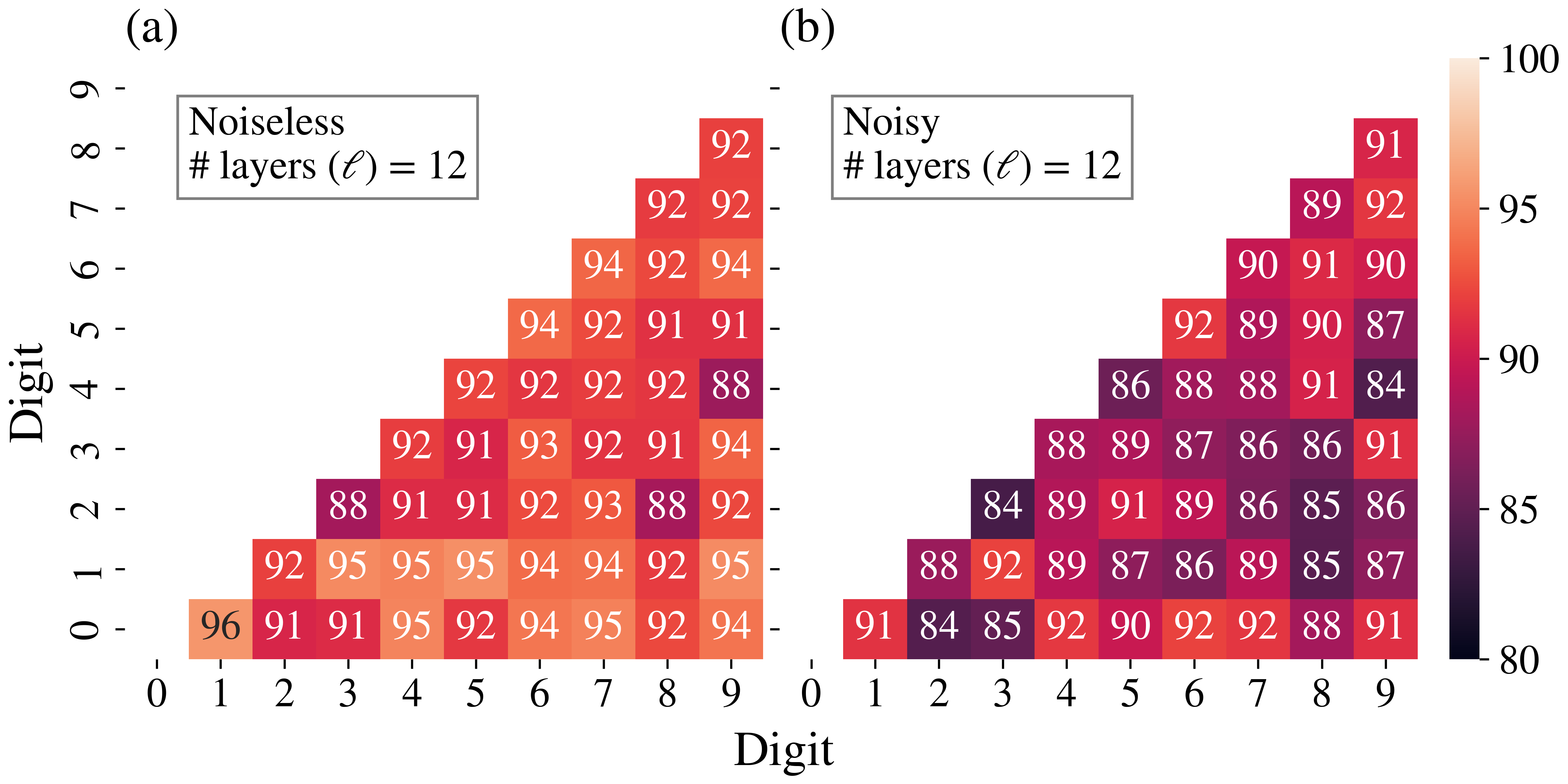}
    \caption{\figupdated{Accuracies for every pair of handwritten digits classified with the digital model, both (a) without and (b) with noise. The number of layers $\ell = 12$ and qubits $n=8$ are fixed for both models.}}
    \label{fig:cnot_compare_diff}
\end{figure*}

\begin{figure}[t!]
    \centering
    \includegraphics[width=0.45\textwidth]{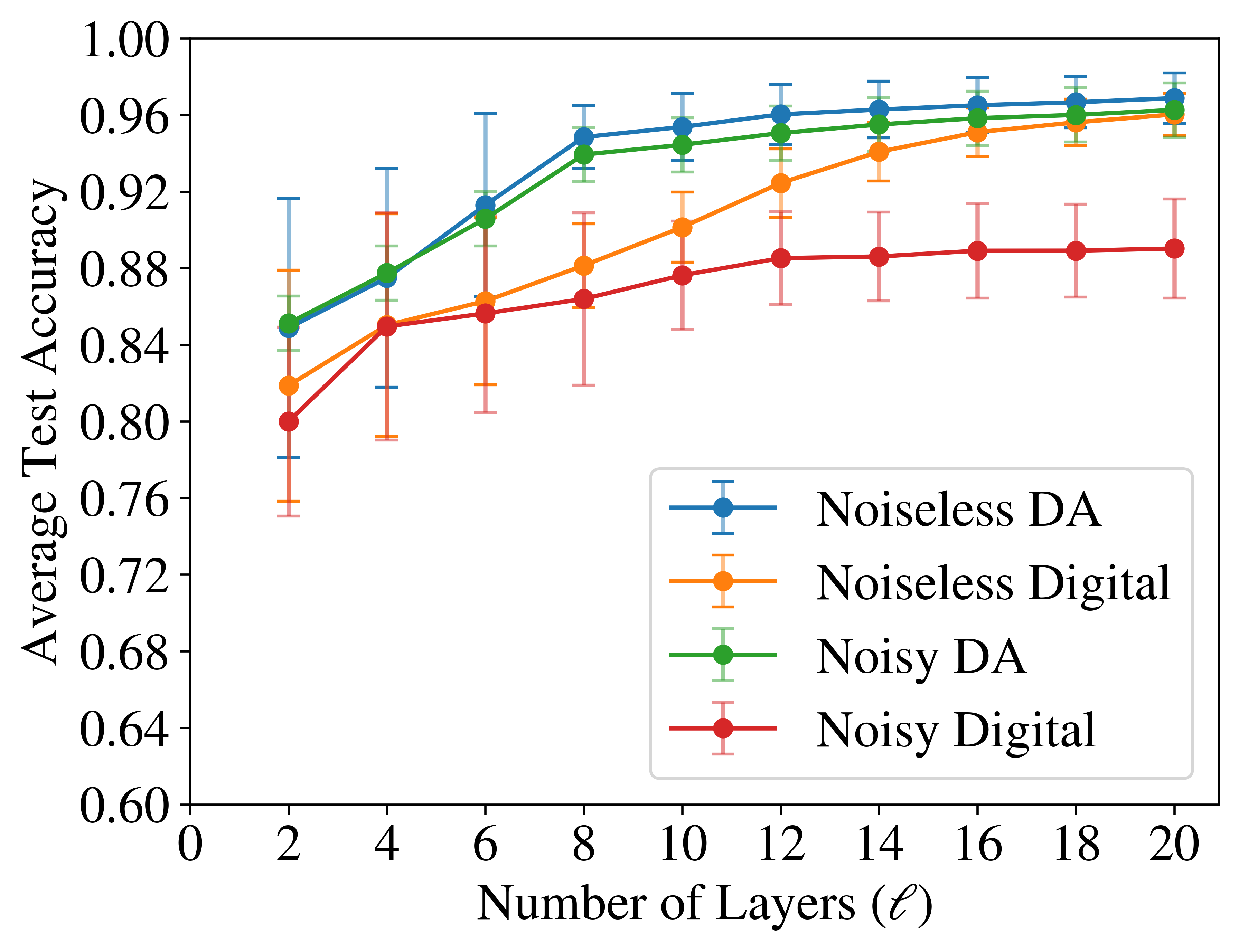}
    \caption{\figupdated{Comparison of classification accuracy, averaged over all pairs of digits, between $R_b/a = 0.87$, $t = 2\pi/\W$, $\D/\W = 0.8$, and $n = 8$ digital-analog (DA) and $n=8$ digital models with their respective noise models. Standard deviation, shown in the error bars, is measured across digit pairs.}}
    \label{fig:compare-depth}
\end{figure}

Fixing the physically optimal digital-analog gate hyperparameters now, we proceed to compare the performance of the digital-analog model to that of the digital model. Figure~\ref{fig:cnot_compare_diff} exhibits the accuracies across every pair of digits under the digital model, both with and without noise. Noiselessly, we found (see Appendix for an example on 1 and 9) that the digital model performs better with increasing depths up to at least $\ell = 20$, but does not improve substantially beyond $\ell = 12$. As we show later in Fig.~\ref{fig:compare-depth}, in the presence of noise, the digital model strictly performs best at $\ell = 12$ due to the noise overpowering the benefit of additional layers beyond 12. We therefore fix $\ell = 12$ for both cases, which place them on equally optimal footing with the digital-analog analysis. Juxtaposing Fig.~\ref{fig:all-digit-compare}(b) and Fig.~\ref{fig:cnot_compare_diff}(a) demonstrates that the digital-analog and digital-models in their optimal hyperparameter regimes produce similar accuracies in the absence of noise. However, juxtaposing Fig.~\ref{fig:all-digit-compare}(e) and Fig.~\ref{fig:cnot_compare_diff}(b) shows that in the presence of noise, the digital model suffers considerably in accuracy, while the digital-analog model has no significant loss. 

This robustness to noise persists across a large range of $\ell$. As shown in Fig.~\ref{fig:compare-depth}, the digital-analog model maintains a consistent advantage in noise robustness over the digital model for $\ell$ up to 20. Moreover, the robustness to noise of the digital model begins to accelerate negatively after $\ell = 12$, due to the effect of the noise on $\Theta(\ell n)$ gates overpowering the expressiveness of additional layers. Meanwhile, the digital-analog model improves, albeit slightly, with increasing $\ell$, suggesting that it requires a substantially larger $\ell$ for noise to overpower learning. This increased depth capability implies that digital-analog models can learn tasks that require more expressivity (i.e., layers) due to their complex nature, extending the near-term utility of VQA's.

In summary, with optimal hyperparameters, the digital-analog system demonstrated a substantial improvement in noise robustness over the digital model. In turn, it allows for improved accuracy in the near term and potential for learning tasks that require larger numbers of layers. A simple explanation for this improvement is the gate fidelity itself. Even for $n=8$ qubits, the fidelity of a digital layer is $\sim 0.918$, while the digital-analog layer has a fidelity of $\sim 0.971$. Importantly, the gate fidelity of the digital-analog layer appears to decrease substantially slower in $n$ than that of the digital layer. Further, the digital-analog model seems to require fewer layers than digital models to achieve similar performance, though the difference in this classification problem is not significant enough to definitively come to such a conclusion.

\subsection{\label{sec:subsec:phase}Boundaries of quantum phases}
We proceed to study a digital-analog learning scheme on quantum data. In this scheme, an unsupervised DA VQA classifies ground states of a chosen Hamiltonian according to their quantum phases. Quantum phases are of fundamental importance in many-body quantum systems. They are of particular relevance in Rydberg atom arrays, in which a plethora of exotic quantum phases have been predicted and discovered~\cite{qpt1,qpt2,qpt3,qpt4,qpt5,qpt6}. To classify phases, we apply an enhanced and appropriately quantized technique common in classical machine learning algorithms known as anomaly detection. In the classical machine learning setting, anomaly detection is commonly used to identify statistical outliers in datasets, with a broad range of applications such as detecting fraudulent bank activity~\cite{hilal2022financial}. Recently, Kottmann et al.~\cite{kottmann2021variational} first considered a quantized version of anomaly detection for the learning of quantum phases. We consider an analogous version built on a Rydberg digital-analog VQA.

Classically, anomaly detection circuits (learning agents) are trained to recognize a specific region of the data space that we consider to be ``normal". This training might be done, for example, by optimizing a neural network to output $0$ on training data sampled from the normal subspace. Effectively, the training serves to overfit a learning agent to a desired region in space. In the testing phase, the learning agent recognizes ``anomalous" data (those outside the normal subspace) by outputting values that are far from zero, thus performing a classification of normality versus anomaly~\cite{nassif2021machine}. If multiple types of anomalies are close in the appropriate metric on the data space, then an anomaly detector can directly distinguish between the anomalies by labeling each as a cluster in data space. In this sense, anomaly detection is a clustering algorithm trained on a single cluster.

Such ideas can be extended to learn quantum information by the following procedure. First, replace the neural network with a Rydberg digital-analog circuit $\CA_d^n$. Next, to ensure the circuit outputs zero on training data, we use the average Rydberg density as the training loss function, given by
\begin{align}
\begin{aligned}
    \CL[\CA(\set{\bth_{ij}})] & = \frac{\bkop{0^{\otimes n}}{\CA^\dag(\set{\bth_{ij}}) \CP_1 \CA(\set{\bth_{ij}})}{0^{\otimes n}}}{n} \\
    & = \frac{1}{n} \sum_{k=1}^n \norm{\braket{1_k}{\psi_{\text{out}}}}^2
\end{aligned}
\label{eq:phase_loss}
\end{align}
where $\ket{\psi_{\text{out}}} = \CA(\set{\bth_{ij}}) \ket{0^{\otimes n}}$ is the circuit output and $\CP_1 = \sum_{k=1}^n \ketbra{1_k}{1_k}$ is the total one-qubit projector onto the Rydberg state. Minimizing Eq.~(\ref{eq:phase_loss}) on the training data ensures that the DA circuit outputs a number very close to 0 in the normal subspace. Moreover, experimental estimation of the average Rydberg density can be done in a number of computational basis measurements independent of $n$, rendering Eq.~(\ref{eq:phase_loss}) a scalable loss function. 

Quantum anomaly detection can be directly extended to quantum phase boundary learning by choosing one phase arbitrarily as the ``normal" phase and letting the rest be anomalies of a different type. Na\"ively, one might expect that each phase would have its own distinct cluster and thus have its own range of circuit outputs. A subtlety, however, is that the circuit may not be sufficiently expressive to have a different value for each phase. Nonetheless, it may still take a dramatically different value when \textit{crossing the boundary} between two phases. As a result, anomaly detection generically can only hope to learn the boundaries of phases rather than quantitatively cluster them. In practice, however, one can visually cluster the phases by qualitatively examining the learned phase diagram.

This picture is still slightly oversimplified because the second-order phase transitions are not perfectly discontinuous jumps but rather continuous transitions with a large gradient (with respect to the Hamiltonian parameters) in the corresponding order parameter. This effect serves to smoothen out the outputs of the digital-analog learning circuit, in particular for finite systems, but does not hinder its efficacy at learning phases; rather, the gradient landscape itself indicates the location of the boundaries.

The quantum data we associate with each point in phase \updated{space} is the ground state of a chosen Hamiltonian $\CH_{\text{learn}}(\bp)$ (generally not the Rydberg Hamiltonian $\CH$) with Hamiltonian parameters $\bp$. We demonstrate that it suffices to carve out the entire phase diagram with a single ground state of a chosen phase. This training dataset size of 1 is particularly appealing for near-term realization because training with large datasets becomes very resource-intensive especially on noisy devices.

We remark that digital-analog circuitry enables an extended picture of phase clustering---in which the quantum data and the learning are performed consecutively in one complete digital-analog circuit---by chaining the digital-analog learning circuit with a step of adiabatic evolution. Adiabatic evolution enables the preparation of approximate ground states at a given phase starting from a ground state of an easy-to-prepare phase (e.g., $\ket{0}^{\otimes n}$) by time-dependent quench dynamics \begin{align}
    \ket{g(\bp)} = e^{- \frac{i}{\hbar} \int_0^1 ds\, \CH_{\text{learn}}[s]} \ket{0}^{\otimes n} 
\end{align}
where $s$ parameterizes a smooth path in phase space---in which every point has a gapped Hamiltonian---such that $\CH_{\text{learn}}[s=0]$ is the Hamiltonian in the easy-to-prepare phase and $\CH_{\text{learn}}[s=1] = \CH_{\text{learn}}(\bp)$. Data preparation can, therefore, be incorporated into the digital-analog circuit as a preliminary analog step before the first 1-qubit layer in Fig.~\ref{fig:DAQC}(a). Note, however, that this analog step is driven according to a Hamiltonian $\CH_{\text{learn}}$ that is generally different the Rydberg Hamiltonian, $\CH$. Adiabatic evolution has been explored extensively not only in Rydberg systems~\cite{qpt5}, but also much more generally as a technique for encoding computations into ground states of local Hamiltonians~\cite{AP2,AP3,AP4,AP5}. Adiabatic evolution produces approximate ground states that lead to a deformed version of the phase diagram that becomes the ground truth for learning~\cite{sahay2022quantum,spinliquid4,lu2022learning}. This unified picture of data preparation and learning further demonstrates the utility of the digital-analog paradigm.

Although adiabatic evolution is necessary for experimental ground state preparation, the manner of ground state preparation is not significant for purposes of algorithm analysis. Consequently, we prepare learning data directly via exact diagonalization in numerical simulation.
\begin{figure*}[ht!]
    \centering
\includegraphics[width=0.9\textwidth]{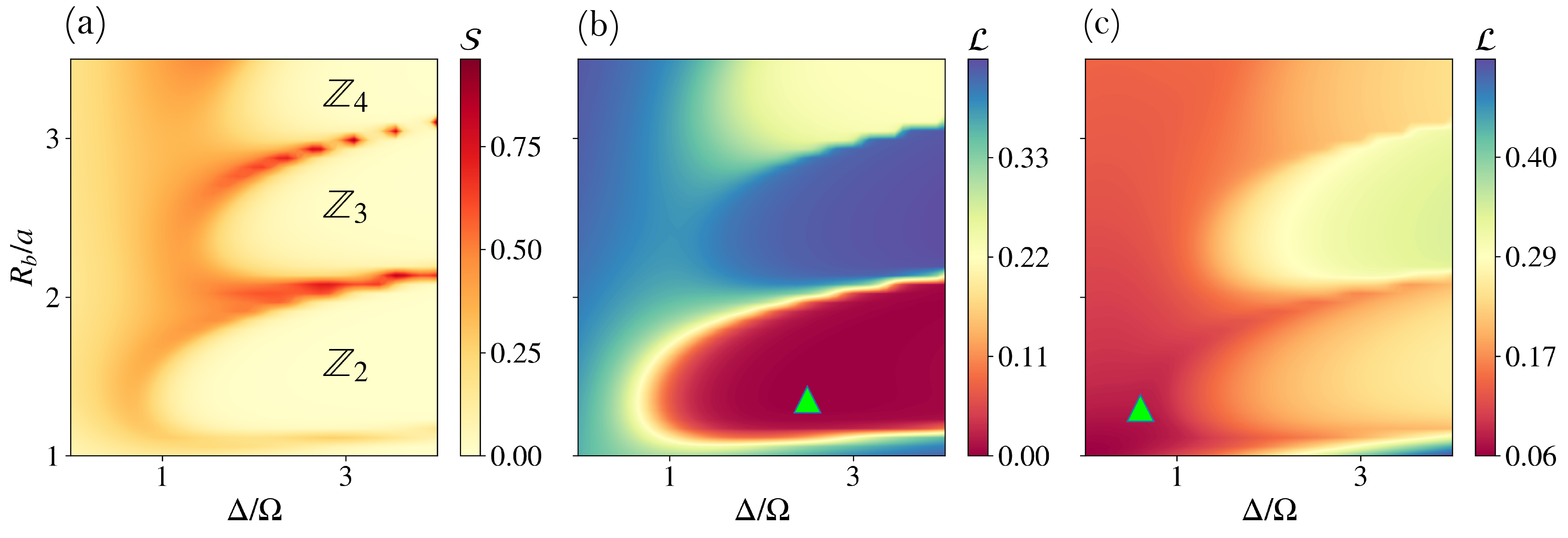}
    \caption{(a): True phase diagram of the 13-qubit Rydberg chain with open boundary conditions, plotted over the phase space of phase parameters $\Delta/\Omega$ and $R_b/a$. Three distinct lobes, referred to as the $\mathbb{Z}_2, \mathbb{Z}_3, \mathbb{Z}_4$ phase regions, are visible. (b): learned phase diagram of a $\ell = 5$ digital-analog circuit $\CA_{\ell = 5}^{n = 13}$. (c): learned phase diagram of a 0-layer circuit (1-qubit gates only). The green triangles mark the training point in the learning procedure.}
    
    \label{fig:ryd_true_learned}
\end{figure*}

The most obvious learning example would be to set $\CH_{\text{learn}} = \CH$, the Rydberg Hamiltonian itself. We first map out a phase diagram classically by using the entanglement entropy \begin{align}
    S[\r_{AB}] = -\operatorname{Tr} [\r_{A} \log \r_{A}] ,
\end{align}
where $\r_{AB}$ is the density matrix of the ground state of the Hamiltonian, partitioned into roughly equal-sized subsystems, and $\r_A$ and $\r_B$ are the reduced density matrices. Figure~\ref{fig:ryd_true_learned}(a) displays the phase diagram on a 13-qubit chain, comprised of the well-known 3 lobes of ordered phases separated by high-entanglement boundaries, surrounded by the disordered phase \cite{Fendley2004, Schachenmayer2010}. Each ordered phase  $\Z_n$ is close in fidelity to the state $\ket{10\cdots 0 1 0 \cdots 0 1 \cdots}$ where each string of 0's is of length $n-1$. 

We apply the anomaly detection VQA on a digital-analog circuit of depth $\ell = 5$, using the same quench hyperparameters as in the digit classification VQA. Figure~\ref{fig:ryd_true_learned}(b) illustrates the output of the digital-analog learning circuit trained on a point in the $\Z_2$ lobe, demonstrating the learnability of the phase diagram. However, this Rydberg Hamiltonian, at least for a small system size, is not a good instance upon which we can analyze the digital-analog VQA because all but the disordered phase are adiabatically connected to their respective computational-basis product states. Even the disordered phase has a relatively low entanglement entropy, being adiabatically connected to the $\ket{+}^{\otimes N}$ state. As such, one might expect that no entanglement in the learning circuit is even necessary to reproduce the phase diagram. This intuition turns out to be correct, as shown in Fig.~\ref{fig:ryd_true_learned}(c). Here, a $\ell = 0$ circuit which has only a single rotation gate per qubit and no entanglement, successfully carves out the phase diagram of the Rydberg chain.

Instead, we keep the one-dimensional spin chain structure of the atoms but consider a 2-body XXZ Hamiltonian with periodic boundary conditions, governed by antiferromagnetic nearest-neighbor (NN) and next-nearest-neighbor (NNN) interactions~\cite{lee2023landau,mudry2019quantum}
\begin{align}
\begin{aligned}
    \CH_{\text{learn}} &= J_3 \sum_r (X_r X_{r+1} + Y_r Y_{r+1}) \\ 
    & \phantom{helloooo} + \a (X_r X_{r+2} + Y_r Y_{r+2}) \\
    & \phantom{hi} +
     J_6 \sum_r [Z_r Z_{r+1} + \a^2 Z_r Z_{r+2}]
\end{aligned}
\end{align}
where, $X, Y, Z$ are the Pauli matrices, $J_3$ and $J_6$ give the strengths of spin-exchange and Ising interactions for nearest-neighbor pairs, respectively. The strength parameter $\alpha$ is defined as $(d_1/d_2)^3$, where $d_1$ is NN distance and $d_2$ is NNN distance. This Hamiltonian cannot be learned by a non-entangling circuit, as shown in Appendix \ref{app:subapp:phase} Fig.~\ref{fig:xyz_rots}.

\begin{figure*}[ht!]
    \centering
    \includegraphics[width=0.7\textwidth]{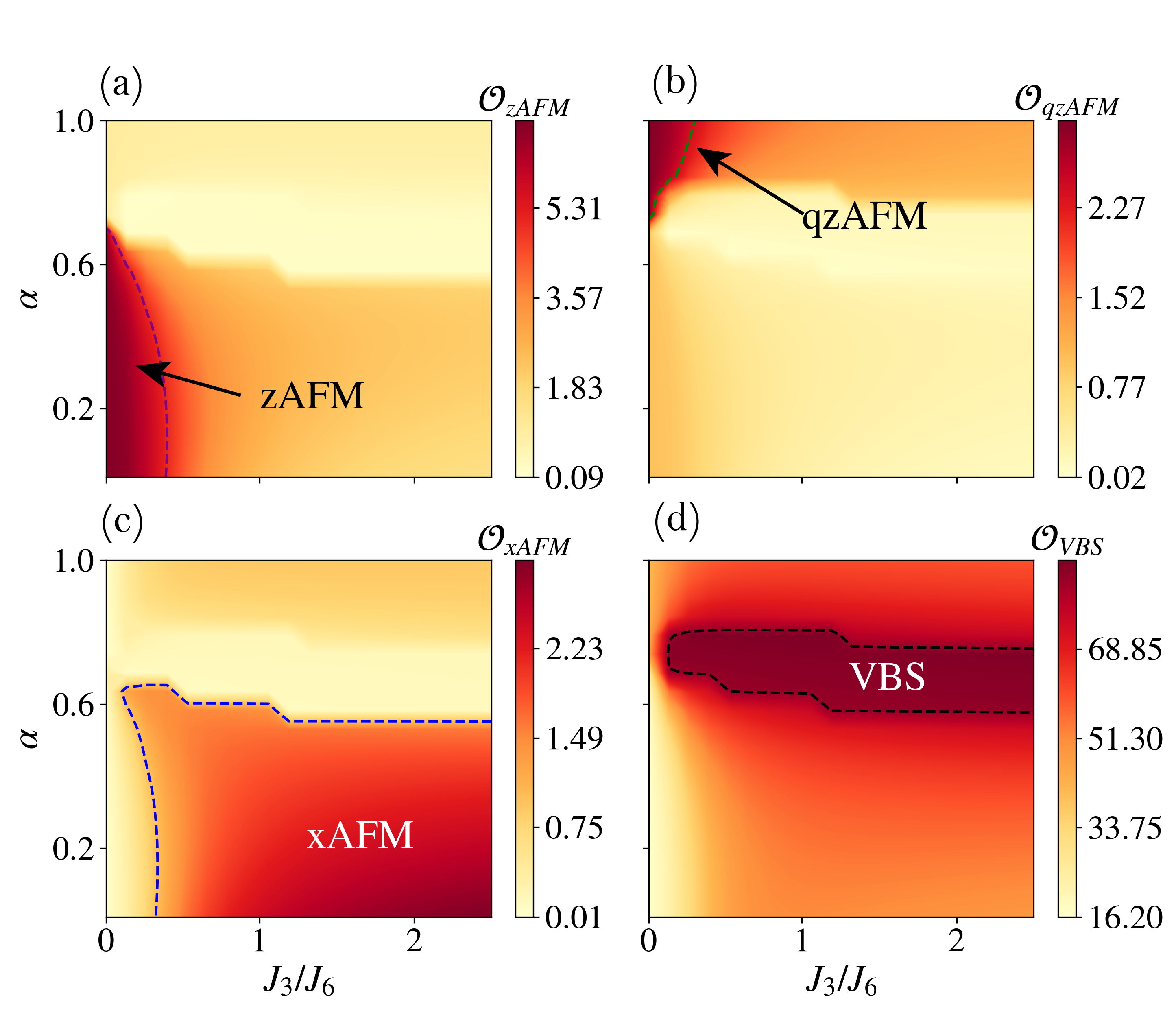}
    \caption{Order parameters of the ground state for the 8-qubit XXZ Hamiltonian with periodic boundary conditions. (a) $\mathcal{O}_{\text{zAFM}}$, the zAFM order parameter (Eq.~\ref{eq:zafm_op}). (b) $\mathcal{O}_{\text{qzAFM}}$, the qzAFM order parameter (Eq.~\ref{eq:qzafm_op}). (c) $\mathcal{O}_{\text{xAFM}}$, the xAFM order parameter (Eq.~\ref{eq:xafm_op}), a proxy for the XY-QLRO phase region. (d) $\mathcal{O}_{\text{VBS}}$, the VBS order parameter (Eq.~\ref{eq:vbs_op}). In each plot, the dashed lines represent the respective phase boundaries, defined for (a)-(d) respectively by contouring around the largest 18\%, 3\%, 50\%, and 9\% values.}
    \label{fig:truepd}
\end{figure*}

Analysis of the XXZ Hamiltonian in the ground state has revealed the existence of four distinct phases, shown in Fig.~\ref{fig:truepd}~\cite{LeeJongYeon, MudryChristopher2019Qptb}. When $J_3 = 0$, the system is strictly classical. In this region, two interactions compete: the NN Ising interactions, which induce antiferrogmagnetic (zAFM) order in the $z$-direction, and the NNN Ising interactions, which induce Quadrupled antiferromagnetic (qzAFM) order. The effect of each interaction is determined by $\a$. The zAFM (Fig.~\ref{fig:truepd}(a)) and qzAFM (Fig.~\ref{fig:truepd}(b)) phases are identified by the two-point $Z$ correlator $C_z(r)$, defined as 
\begin{align}
    C_z(r) = \langle \psi | Z_0 Z_r | \psi \rangle 
\end{align}
where $r$ indexes the qubit, $r=0$ represents the chosen ``central" site in the chain due to periodic boundary conditions (here, $r = 2$), and $\psi$ is the ground state of the Hamiltonian.

The zAFM and qzAFM order parameters are defined from this correlator as 
\begin{align}
    \mathcal{O}_{\text{zAFM}} & = \sum_{r=1}^{n} e^{i \pi r} C_z(r)
    \label{eq:zafm_op} , \\
    \mathcal{O}_{\text{qzAFM}} & = \sum_{r=1}^{n} e^{i \pi r/2} C_z(r)
    \label{eq:qzafm_op} ,
\end{align}
where $n$ the number of sites in the chain.

For $\a \sim 0$, the zAFM order dominates when $J_3 < J_6$, while a symmetric gapless XY phase with quasi-long-range order (XY-QLRO) prevails when $J_3 > J_6$ (Fig.~\ref{fig:truepd}(c)). This quantum, highly entangled phase does not have a local order parameter, but can be qualitatively inferred by the xAFM order parameter that decays algebraically in the phase, and is given by
\begin{align}
    \mathcal{O}_{\text{xAFM}} = \sum_{r=1}^{n} e^{i \pi r} C_x(r)
    \label{eq:xafm_op}
\end{align}
where 
\begin{align}
    C_x(r) = \langle \psi | X_0 X_r | \psi \rangle.
\end{align}
is the two-point $X$ correlator. 

The final phase region occurs roughly where $\alpha \approx 0.7$ and $J_3/J_6 \to \infty$. This phase (Fig.~\ref{fig:truepd}(d)) is referred to as valence bond solid (VBS), which consists of dimerized patterns of spin singlets.  The VBS phase is characterized by the two-point correlator
\begin{align}
\begin{aligned}
    C_{\text{VBS}}(r) & =  \langle \psi | (\vec{\s}_1 \cdot \vec{\s}_0 - \vec{\s}_0 \cdot \vec{\s}_{-1}) \\ 
    & \phantom{hellooo} \times (\vec{\s}_{r+1} \cdot \vec{\s}_r - \vec{\s}_r \cdot \vec{\s}_{r-1})|\psi \rangle
\end{aligned}
\end{align}
where $\vec{\s}_{i} = (X_i, Y_i, Z_i)$ is the Pauli spin matrix vector.
Likewise, the VBS order parameter is defined as
\begin{align}
    \mathcal{O}_{\text{VBS}} = \sum_{r=1}^{n} e^{i \pi r}C_{\text{VBS}}(r) .
    \label{eq:vbs_op}
\end{align}
These order parameters are by symmetry real. In contrast to Rydberg density wave phases, VBS and QLRO cannot be adiabatically connected to product states and are entangled. The same holds for zAFM and qzAFM in a perfect system with periodic boundaries. For example, under such conditions the ground state of the zAFM is a cat state $\propto \ket{0101\dots} + \ket{1010\dots}$. 
\begin{figure*}[ht!]
    \centering
    \includegraphics[width=0.7\textwidth]{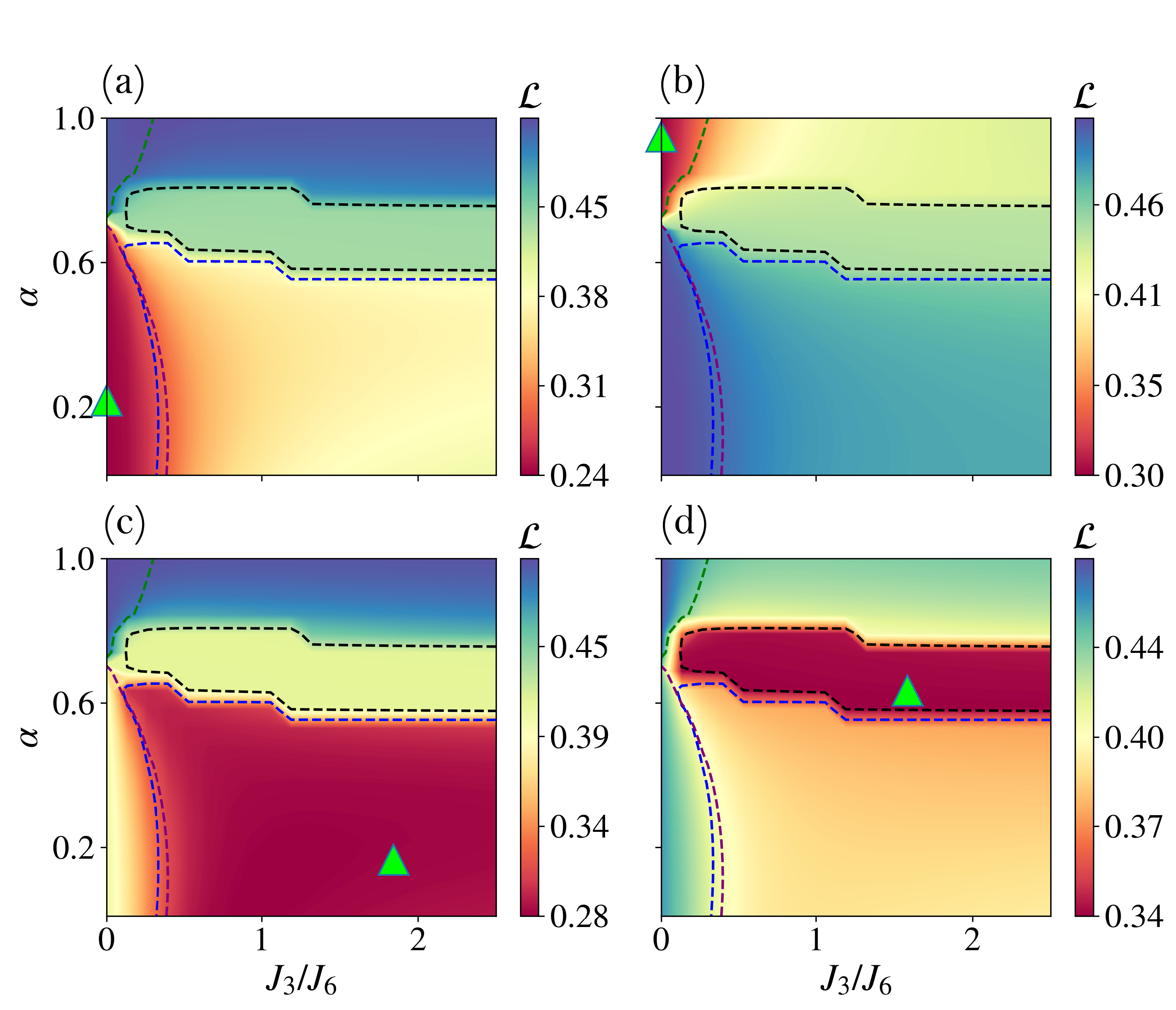}
    \caption{Four learned phase diagrams for the 8-qubit XXZ Hamiltonian with parameters $\a$ and $J_3/J_6$, achieved on a digital-analog circuit $\CA_{\ell = 2}^{n = 8}$ without noise. The green triangles indicate the point of training.
    The dashed contours outlining each phase are the same as in Fig.~\ref{fig:truepd}, demonstrating remarkably that the digital-analog circuit reproduces all of the boundaries specified by the four order parameters at once for each training area.}
    \label{fig:XXZ_LPD_4x4}
\end{figure*}

Figure~\ref{fig:XXZ_LPD_4x4}(a)-(d) depicts the output of digital-analog anomaly detection on the 8-qubit XXZ chain, with four different training points chosen each in one phase region. Although the circuits were simulated noiselessly here, we observe already two noteworthy results. First, generically to digital-analog learning, the successful demarcation of phases alone is a nontrivial result, because we used the same physically justified hyperparameters as in the noiseless digit classification algorithm. This result provides stronger evidence that the quench dynamics parameters can be optimally chosen without tailoring them to a specific problem, and support the validity of our physical justifications. It moreover provides an explicit demonstration of the versatility of the analog layer as a tool for entanglement generation. Second, regardless of which phase acted as the training point, each learning outcome produced demarcations of the entire phase diagram, which \updated{if} relying on order parameters alone would have required examining multiple order parameters simultaneously to deduce. In this sense, anomaly detection may have additional utility for the study of Hamiltonians whose phases are not visible by a single global order parameter.

In the absence of noise, there does not appear to be a substantial advantage in the number of layers $\ell$ required for successful learning between digital-analog and digital circuitry. This lack of advantage is not evidence against a general advantage for digital-analog circuitry, however, because phase boundaries of the XXZ Hamiltonian simply appear to be learnable with relatively low entanglement in the circuit (see Fig.~\ref{fig:depth2_noisycomp} in Appendix \ref{app:subapp:phase}). Ground states of local one-dimensional Hamiltonians generally have either constant entanglement entropy for gapped phases or $\log N$ for gapless phases, which limits the number of layers needed to distinguish phases. It is likely that with even more complex Hamiltonians, higher dimensions, or dynamical phase diagrams with non-area law states, a layer advantage may emerge.

We next apply the noise models discussed in Section~\ref{sec:rydberg} for purposes of a digital-analog and digital circuit comparison. To obtain a quantitative metric of performance akin to accuracy in the digit classification problem, we define the sharpness of a learned phase diagram as \begin{align}
    \s_{\text{learn}}(\set{\bth_{ij}}) = \operatorname{SD}_{\bp} \left(\norm{\nabla_{\bp} \CL[\CA^n_\ell(\set{\bth_{ij}})]}^2 \right)
    \label{eq:sharpness}
\end{align}
where $\operatorname{SD}_{\bp}$ is the standard deviation over the uniform distribution of the two-dimensional parameter space $(J_3/J_6, \a)$. Intuitively, the sharpness of a learned phase diagram measures how distinctly the boundaries can be identified. A well-trained circuit that outputs nearly gradient-free interiors and rapid-transitioning boundaries produces very large $\s_{\text{learn}}$. By contrast, a poorly trained circuit with less distinct, smoother transitions results in lower sharpness and less informative learning results. In practice, $\s_{\text{learn}}$ is estimated by calculating numerical gradients and sample standard deviations over a discretized uniform mesh grid of the parameter space. We note that circuit noise impacts sharpness both by reducing the ability of the circuit to distinguish phases, thereby decreasing $\s_{\text{learn}}$, and introducing small local fluctuations that increase $\s_{\text{learn}}$. However, the latter effect is by definition substantially smaller than the first and can be thought of as negligible, so that $\s_{\text{learn}}$ remains a well-defined measure of noise robustness in phase boundary learning. Moreover, noise-based effects on $\s_{\text{learn}}$ are mitigated by averaging over multiple trials and, therefore, primarily affect the variance of $\s_{\text{learn}}$.

\begin{comment}
Originally, this figure was labeled_sharpness_comp.png
\end{comment}
\begin{figure}[t!]
    \centering
    \includegraphics[scale=0.07]{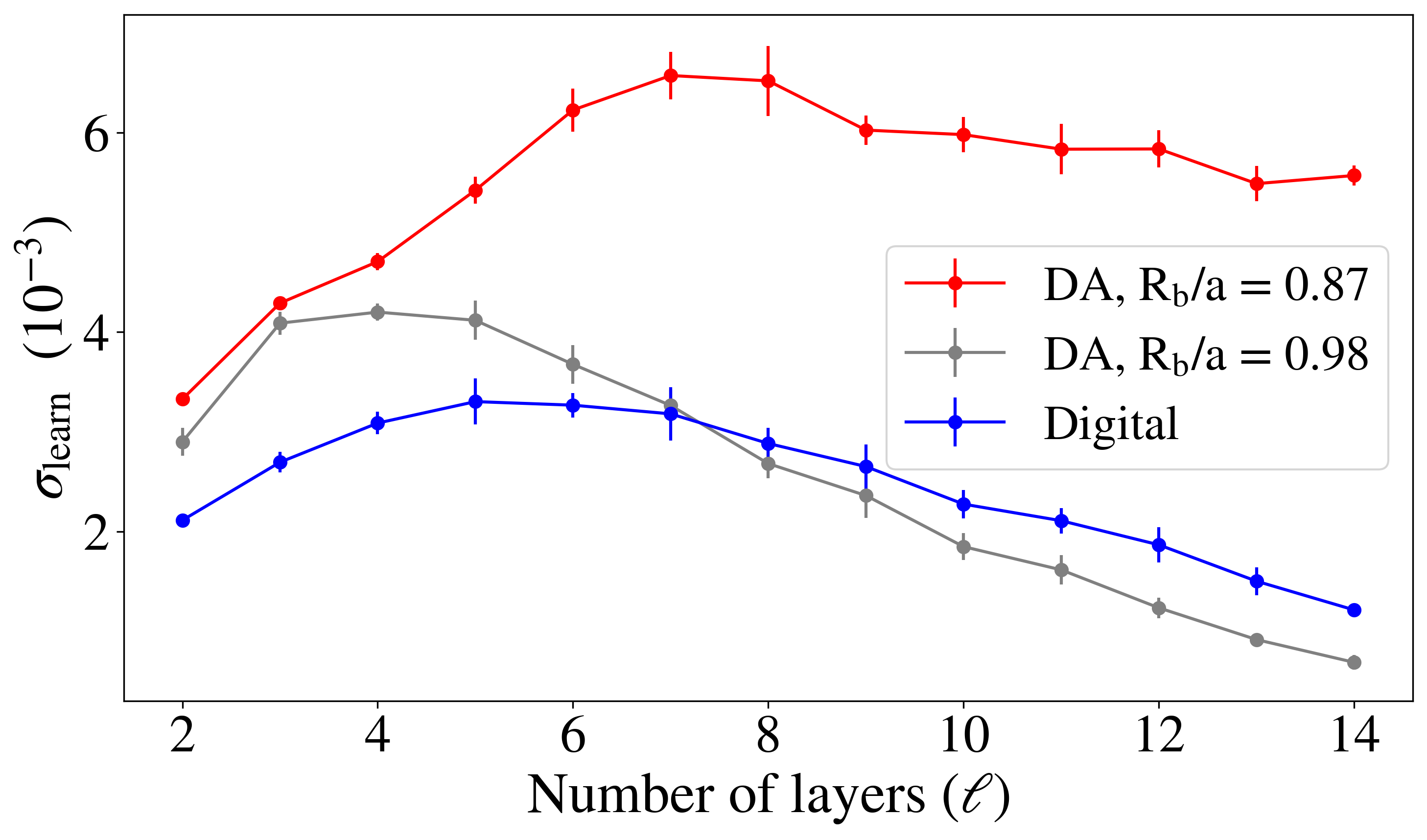}
    \caption{\figupdated{Comparison of $\s_{\text{learn}}$ estimated numerically as a function of $\ell$ on digital-analog $\CA^{n=8}_{\ell}$ and digital $\mathcal{D}^{n=8}_{\ell}$ circuits for the XXZ Hamiltonian. Each circuit is trained on a zAFM point. Error bars denote 1 standard deviation over randomness in the noise. The advantage in sharpness is visible for a good hyperparameter $R_b/a = 0.87$, but does not persist when the choice of hyperparameter $R_b/a = 0.98$ is bad.}}
    \label{fig:sharpness_comp}
\end{figure}

\begin{comment}
    original figure was labeled_noisy_sidebyside.png
\end{comment}

\begin{figure}[ht!]
    \centering
    \includegraphics[width=0.4\textwidth]{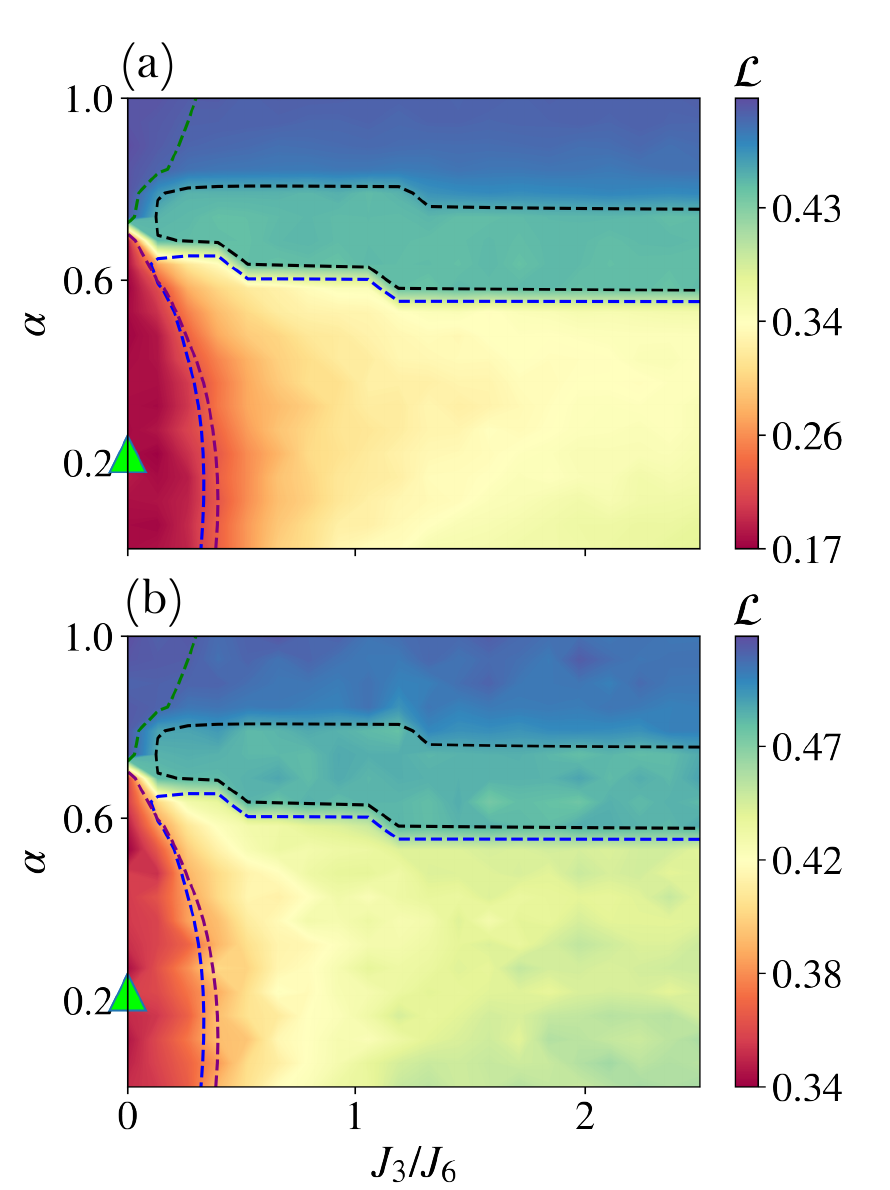}
    \caption{\figupdated{Learned phase diagrams of the XXZ Hamiltonian with respect to parameters $\a$ and $J_3/J_6$ for (a) digital-analog and (b) digital circuits with $n = 8, \ell = 14$ in the presence of their respective noise models. Dotted contours overlay the true phase boundaries.}}
    \label{fig:noisy_comp_d14}
\end{figure}

In Fig.~\ref{fig:sharpness_comp}, we calculate $\s_{\text{learn}}$ numerically on learned phased diagrams averaged over 20 independently trained learning circuits, again with $n = 8$ but with the training point fixed in the zAFM phase for consistency. We repeat this procedure 5 times to provide a simple error bound, shown in the error bars. We observe that when the hyperparameter choice is good ($R_b/a = 0.87$)---in terms of the analysis in Section~\ref{hyperparameters}---the sharpness improves at larger depths for digital-analog circuits, whereas the noise quickly reduces the sharpness of the digital circuits. \updated{W}e are primarily interested in the large-$\ell$ behavior as it indicates the noise robustness for learning problems that require greater circuit depths, e.g., 2D ground states or dynamical phases (see next section). Consequently, digital-analog circuits appear also to be more robust than digital circuits in learning phase boundaries, in agreement with findings from digit classification. We observe, however, that this advantage is lost if the choice of hyperparameter with respect to noise is poor, as shown by the grey line ($R_b/a = 0.98$). Hence, we can also explicitly see in phase boundary learning that the choice of hyperparameter makes a substantial impact on the learning advantage of digital-analog circuits. We observe visual differences in sharpness at $\ell = 14$ between digital-analog and digital circuitry in Fig.~\ref{fig:noisy_comp_d14}\updated{, namely blurrier phase boundaries and more artifacts in the digital case.  Notably, the performance of the digital circuit with $\phi = \pi/4$ is \textit{qualitatively} worse. In Fig.~\ref{fig:bad_cnot_phase} of Appendix~\ref{app:calculations}, we show that a digital $\phi = \pi/4$ circuit is unable to identify the VBS phase at all}.

\section{\label{sec:conclusion}Conclusion \& Outlook}
We have observed numerically that both in digit classification and phase boundary learning, digital-analog circuits achieve substantial practical advantages over their digital counterparts as the quantum agents in the VQA. In both cases, the digital-analog circuit demonstrated a substantially greater robustness to noise. The \updated{comparatively larger} gate fidelity of the \updated{digital-analog} entangling layer \updated{provided a simple explanation of} this improvement. A subtlety of the error advantage was that it depended on the choice of $R_b/a$, based on the physical tradeoff of large separations ($R_b/a \ll 1$) leading to relatively smaller position noise effects and small separations ($R_b/a \sim 1$) leading to greater interaction strength. Properly trading off these effects by choice of $R_b/a = 0.87$ enabled the noise advantage. Similarly, choices of $t = 2\pi/\W$ led to optimal performance of the digital-analog circuits. While seemingly complicated, the hyperparameter choices have physical justifications which allow them to be more easily understood, providing an advantage over digital gates which do not have any such parameters that can be tuned to physical optimality. Moreover, our work provides evidence that these choices are indeed roughly independent of the learning problem, since the same choices led to comparatively advantageous performance of digital-analog circuits in both problems we investigated. This universality is expected, since the parameters are inherently related to the physics of the circuit and not the problem it learns. Consequently, no re-derivation of these hyperparameters is necessary from one learning problem to the next.

We showed also that for a problem in which circuits of a few layers $\ell$ do not perform well on, digital-analog circuitry enables learning to comparable accuracies in substantially smaller $\ell$. We reiterate that the fact that digital-analog circuits can learn at all is itself notable. Practically, even without the digital-analog advantages we have shown, it is much easier to realize analog learning because it requires much fewer local control on qubits and acts as one global operation rather than $\Theta(n)$ local, non-commuting operations. Theoretically, our results suggest that certain quench times and Hamiltonian parameters are admissible to learning while the rest are not, raising questions that we discuss further below about the ranges of good hyperparameters.

% For purposes of practical implementation, it is useful to examine whether phase boundaries can still be learned under global rotation parameters, i.e. in each layer all qubits are rotated by the same angle. Experimental realization aside, a 

A useful future direction would involve putting the physical justifications in this work on more rigorous theoretical grounds. Specifically, it is an interesting question as to the general robustness of the optimal evolution time $t = 2\pi/\W$. We justified this choice by arguing that (a) sufficiently small times would not offer enough entanglement, (b) sufficiently large times are known to generically lead to thermalization, and (c) $2\pi/\W$ is the natural time scale of the system. Yet this justification accounts only for the small- and large-time limits, and does not enable us to understand what range of intermediate times are generally useful. A similar optimality question concerns an analytical solution and more rigorous physical understanding of the tradeoff effect between interaction strength and gate fidelity when tuning $R_b/a$. \updated{Similarly, with either greater computational resources or more efficient noise simulation algorithms, a related direction of extension consists of expanding the noise model to include incoherent sources.}

Another potential direction involves deforming part of the phase boundary learning problem to examine its versatility. One possibility is to study adiabatically prepared states for their own utility. While we discussed the use of adiabatic evolution to generate approximate ground states that reproduce our phase diagrams, it is in general infeasible to do so in polynomial time because the preparation of ground states is computationally hard. A classic result in quantum complexity, for example, states that even determining whether the ground state energy of a local Hamiltonian is below a certain threshold is \textbf{QMA}-hard \cite{Schuch2009}, where \textbf{QMA} is a quantum analog of the classical complexity class \textbf{NP}. Any adiabatically prepared ensemble of states produces quantum state classification task which we can learn, though it may not correspond to a phase diagram. It is an interesting question as to what physics we may learn from this class of quantum classification problems. 

Another option for learning the ground-state phase diagram in particularly hard instances of the Hamiltonian may be to turn to quench dynamics themselves. Consider $\ket{0}$ quenched for some intermediate time long enough to leave a ``signature" of the phase behind but not long enough for thermalization and quantum chaos to ensue. Whether or not such intermediate-time signatures even always exist is an open question, but recent studies on spin chains have demonstrated promising results~\cite{haldar2021signatures}. Assuming the presence of intermediate-time quench phase signatures, an intriguing extension of quantum phase learning would be whether and how anomaly detectors could learn the signatures without supervision. Even without signatures, the use of intermediate-time quenched states may be useful for QML predictions of dynamical phase transitions, although this may require input from states quenched with many different quench times.

\section*{\label{sec:acknolwedgements}Acknowledgements}
The authors thank Jonathan Wurtz for an insightful discussion about practical error models on Rydberg systems.
Authors affiliated with QuEra Computing Inc.~acknowledge the support by the DARPA-STTR award (Award No.~140D0422C0035). S.F.Y.~thanks the NSF through the CUA PFC and the Qu-IDEAS HDRI for funding. H.Y.H.~is grateful for the support from the Harvard Quantum Initiative. The computations in this paper were run on the FASRC Cannon cluster supported by the FAS Division of Science Research Computing Group at Harvard University.

\bibliography{apssamp}% Produces the bibliography via BibTeX.

\clearpage
\newpage

\appendix
\section{\label{app:experiment}Experimental Considerations}

In this Appendix, we briefly discuss an experimental technique for the implementation of a digital-analog system on Rydberg atom arrays. One such realization relies on two-photon coupling processes. A complete discussion of these experimental considerations can be found in Bluvstein et al.~\cite{bluvstein2022quantum}.

\begin{figure}
    \centering
    \includegraphics[scale=0.5]{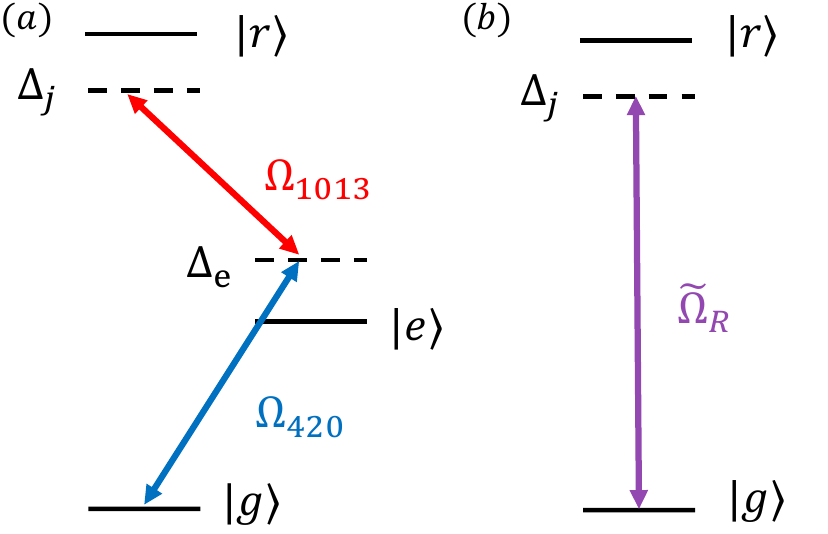}
    \caption{Experimental implementation protocol of the analog two-level system. (a) Two-photon process couples the ground state of $^{87}\text{Rb}$ atoms prepared in the hyperfine and magnetic sublevel $\ket{g}=\ket{5S_{1/2},F=2,m_{F}=2}$ to a Rydberg state $\ket{r}=\ket{70S_{1/2},m_{J}=1/2}$ via an intermediate state $\ket{e}=\ket{6P_{3/2},F=3,m_{F}=3}$ with detuning $\Delta_{e}$ with single-photon Rabi frequencies $\Omega_{420}$ and $\Omega_{1013}$ respectively. States are chosen to maximize Rabi frequency. (b) Effective two-level system $\left(\ket{g}\leftrightarrow \ket{r}\right)$, with Rabi frequency $\Tilde{\Omega}_{R} = \Omega_{420}\Omega_{1013}/2\Delta_{e}$ and detuning $\Delta_{j}.$}
    \label{fig:analog_implementation}
\end{figure}

\begin{figure}
    \centering
    \includegraphics[scale=0.5]{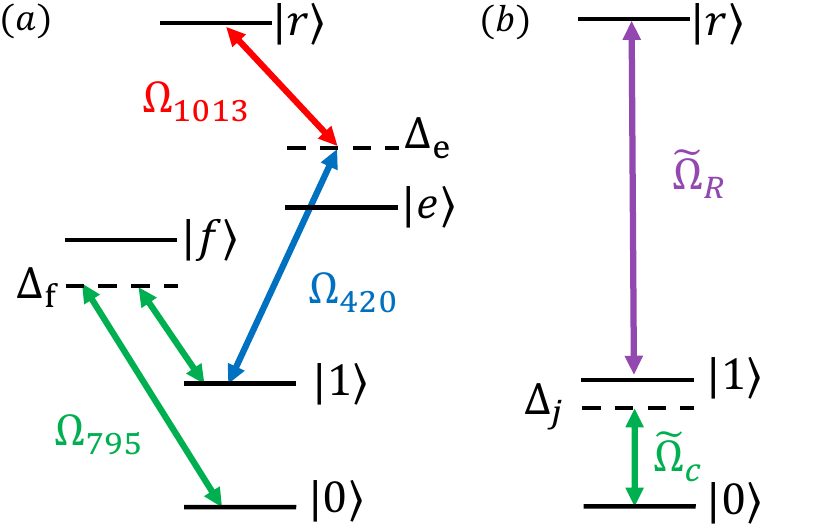}
    \caption{Implementation protocol for a hybrid digital-analog Rydberg system. (a) Besides the Rydberg two-photon process in the analog Hamiltonian simulator, a secondary two-photon process couples the two ground state manifolds of $^{87}\text{Rb}$ namely $\ket{0}=\ket{5S_{1/2},F=2,m_{F}=0}$ to $\ket{1}=\ket{5S_{1/2},F=1,m_{F}=0}$ via an intermediate state $\ket{f}=\ket{5P_{1/2},F=2,m_{F}=-1}$ with a far off resonance detuning $\Delta_{f}$ using circularly polarized $795$-nm lasers with single-photon Rabi frequencies $\Omega_{795}$. (b) Effective three-level ladder system coupling $\ket{0}\leftrightarrow \ket{1}$ and $\ket{1}\leftrightarrow \ket{r}$, with Rabi frequencies  $\Tilde{\Omega}_{c} = \Omega_{795}^{2}/2\Delta_{f}$ and $\Tilde{\Omega}_{R} = \Omega_{420}\Omega_{1013}/2\Delta_{e}$ and detuning $\Delta_{j}$ respectively.}
    \label{fig:digial_implementation}
\end{figure}

Figures \ref{fig:analog_implementation} and \ref{fig:digial_implementation} highlight the key levels in the atomic structure in $^{87}\text{Rb}$. In the analog-only configuration, atoms are prepared in a specific ground state of the $5S_{1/2}$ manifold defined by the hyperfine state and magnetic sublevel $\ket{g}=\ket{F=2,m_{F}=2}$. Atoms interact by coupling them to the $70S_{1/2}$ Rydberg state of the magnetic sublevel $\ket{r} = \ket{m_{J}=1/2}$ via a two-photon scheme with 420 nm $\sigma^{+}$-polarized light and 1013 nm $\sigma^{-}$-polarized light. These states are chosen to maximize Rabi frequency at fixed laser power. In the digital implementation, qubit states are defined by the clock states in the ground state manifold $\ket{0}=\ket{F=1,m_{F}=0}$ and $\ket{1}=\ket{F=2,m_{F}=0}$ to minimize magnetic field sensitivity. Qubit rotations are performed by a two-photon Raman process using 795 nm $\sigma{+}$ light that couples $\ket{0}$ and $\ket{1}$. Interactions between qubits are enabled by a two-photon transition that couples state $\ket{1}$ to the Rydberg state $\ket{r}$. Therefore, it is possible to program a series of pulses to enable digital-analog control over the qubits. 

\section{\label{app:calculations}The Generalized CNOT Gate}
\updated{
In the main text, we described a controlled-NOT gate by a $\pi/4$-time evolution under a two-qubit Hamiltonian $e^{-i \frac{\pi}{4} (I_1 - Z_1) (I_2 - X_2)}$. We here derive this expression. $(I_1 - Z_1) (I_2 - X_2) = I - Z_1 - X_2 + Z_1 X_2$. Since all these terms commute, we may factor the exponential as \begin{align}
e^{-i \frac{\pi}{4} (I_1 - Z_1) (I_2 - X_2)} & = e^{-i \frac{\pi}{4}} e^{i \frac{\pi}{4} Z_1} e^{i \frac{\pi}{4} X_2} e^{-i \frac{\pi}{4} Z_1 X_2} .
\end{align}
Note that if $A^2 = I$, then $e^{i \th A} = \cos(\th) A + i \sin(\th) A$, and thus the above resolves to \begin{align}
    e^{-i \frac{\pi}{4}} \frac{1}{\sqrt{2}} (I + i Z_1) \frac{1}{\sqrt{2}} (I + i X_2) \frac{1}{\sqrt{2}} (I + i Z_1 X_2) .
\end{align}
We can readily check that the above matrix is the CNOT gate \begin{align}
    \text{CX}_{1 \to 2} = \begin{pmatrix}
        1 & 0 & 0 & 0 \\
        0 & 1 & 0 & 0 \\
        0 & 0 &  0 & 1 \\
        0 & 0 & 1 & 0 
    \end{pmatrix} .
\end{align}
The generalized CNOT gate is given by $\text{CX}_{1 \to 2}(\phi) = e^{-i \phi (I_1 - Z_1) (I_2 - X_2)}$.
Using the methods in the main text of our paper, we vary $\phi \in [0, \pi/4]$ and compute the accuracy in classifying digits. Figure~\ref{fig:optimal_digital_digit}
demonstrates that the optimal performance occurs approximately halfway in the range, giving an optimal choice of $\phi = \pi/8$. \begin{figure}
    \centering
    \includegraphics[width=0.9\linewidth]{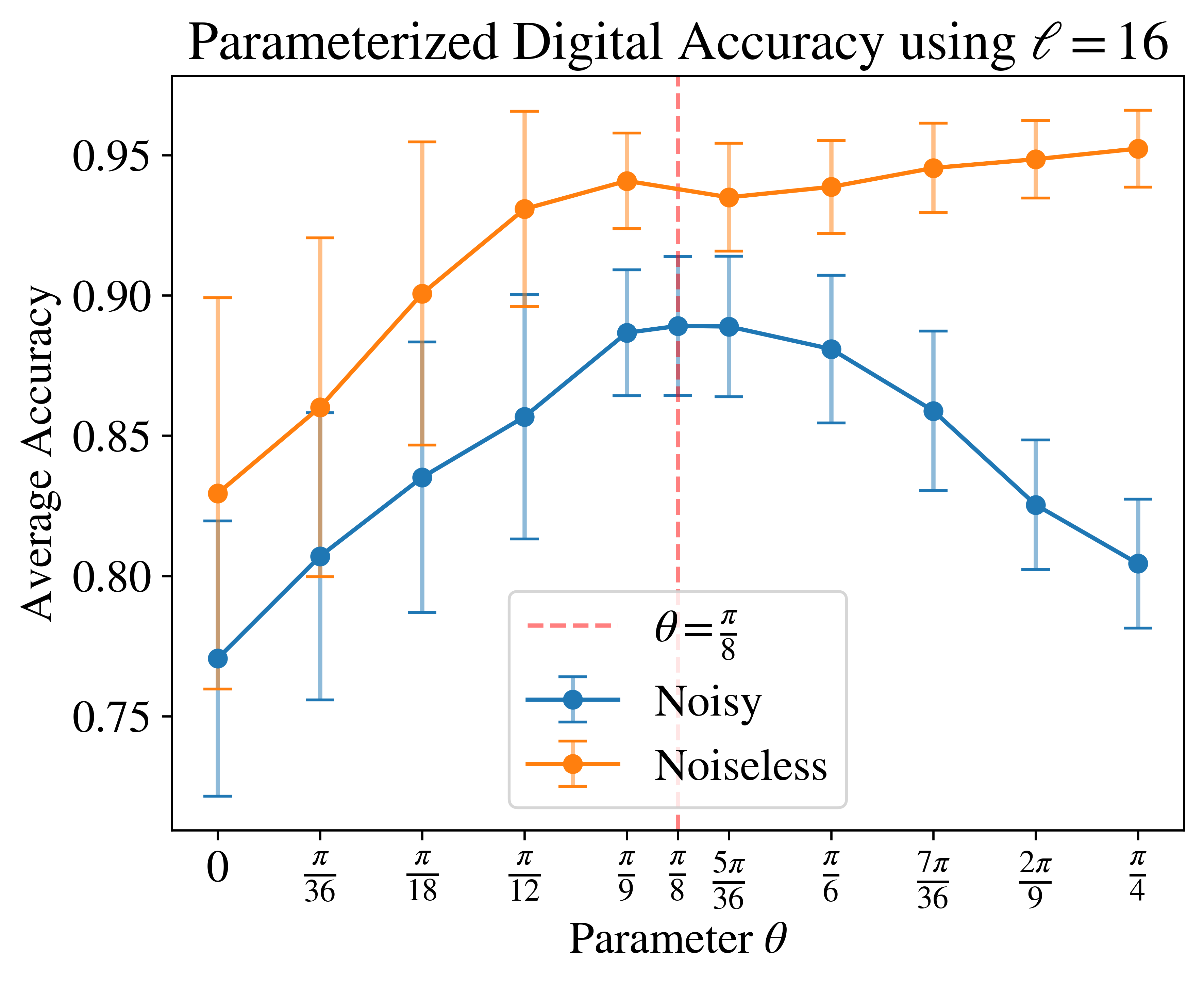}
    \caption{\figupdated{
    Average accuracies over digit comparisons of the parameterized digital circuits on the MNIST dataset with varying the parameter $\theta$, while keeping the number of layers constant at $\ell = 16$. The vertical dashed line denotes the optimal point for the noisy circuit at $\theta = \pi/8$.
    }}
\label{fig:optimal_digital_digit}
\end{figure}
We observe that the performance difference relative to the simple choice of $\pi/4$ is substantial, with an accuracy gap of 8.5\%. This gap holds true even more significantly for phase boundary learning. Figure~\ref{fig:bad_cnot_phase} demonstrates as an example, a stark difference in the learned phase diagrams of the XXZ Hamiltonian between $\pi/4$ and $\pi/8$ (shown in Fig.~\ref{fig:noisy_comp_d14} of the main text). In the $\pi/4$ case, the VBS phase is not identified at all.
\begin{figure}
    \centering
    \includegraphics[width=0.45\textwidth]{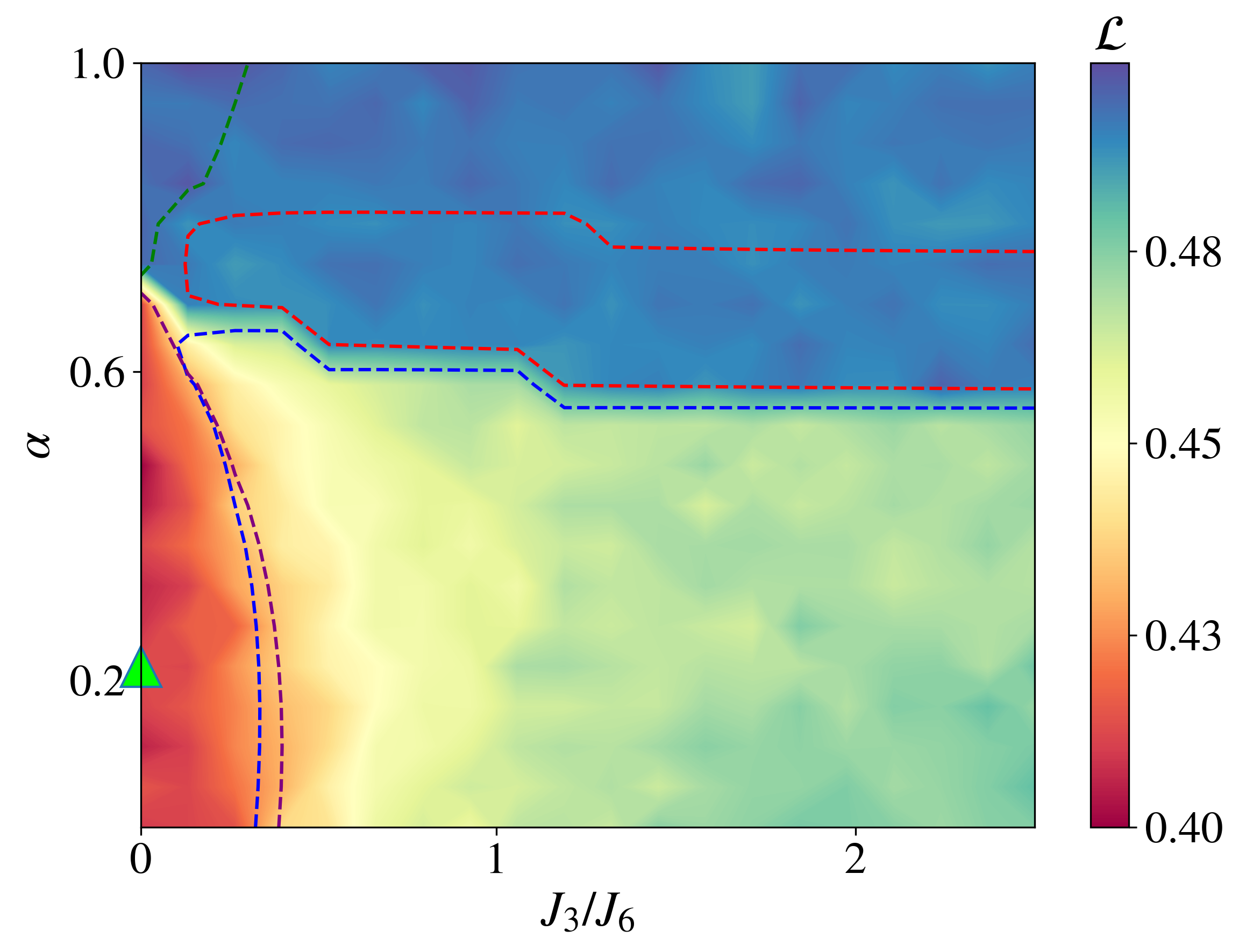}
    \caption{\figupdated{Learned phase diagram of the XXZ Hamiltonian with respect to parameters $\a$ and $J_3/J_6$ for a digital ($\phi = \pi/4$) circuit with $n = 8, \ell = 14$ in the presence of noise. Dotted contours overlay the true phase boundaries. Notably, the VBS phase (red contour) is not learned at all.}}
    \label{fig:bad_cnot_phase}
\end{figure}
These differences highlight that a choice of $\pi/4$ over $\pi/8$ causes significant, even qualitative, losses in QML performance. This gap strengthens the separation between digital and digital-analog schemes, as in some devices the choice of $\pi/8$ might need to be built out of more native gates, which could amplify noise and worsen performance. Yet if $\pi/4$ is chosen, the performance will worsen anyway. By contrast, the digital-analog scheme avoids these implementation subtleties due to its native implementation on neutral atom devices.
}

\section{\label{app:further_discussion}Further Discussion}

\subsection{\label{app:subapp:MNIST}Binary classification}

    \subsubsection{Additional hyperparameter search}
    \begin{figure*}
        \centering
        \includegraphics[width=0.75\textwidth]{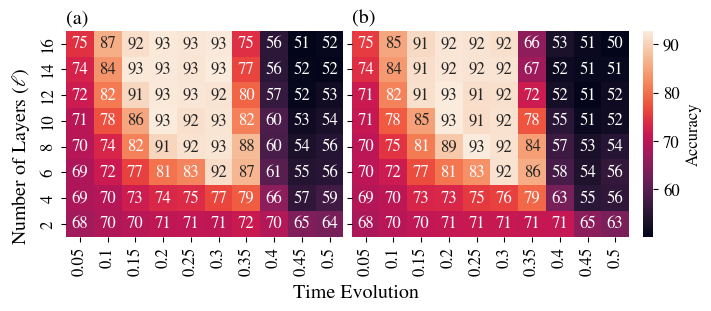}
        \caption{Hyperparameter grid search of the time evolution constant and the number of layers in the circuit on the comparison between 1 versus 9 for (a) the noiseless model and (b) the noisy model. All models are using $R_b/a = 0.87$ and $n = 8$.}
        \label{fig:087_1v9}
    \end{figure*}
    
    The choice for the time evolution constant may seem suboptimal when looking solely at the 3 versus 8 digit comparison, but bringing our attention to the 1 versus 9 digit comparison, we can see that the optimal region is much more constrained. 

    In Fig.~\ref{fig:087_1v9}, we see that the accuracy drops after 0.3 for the time evolution constant. This indicates we should pick our time evolution parameter to be more deeply in the optimal region, especially for robustness to noise. 

    \begin{figure}
        \centering        \includegraphics[width=0.45\textwidth]{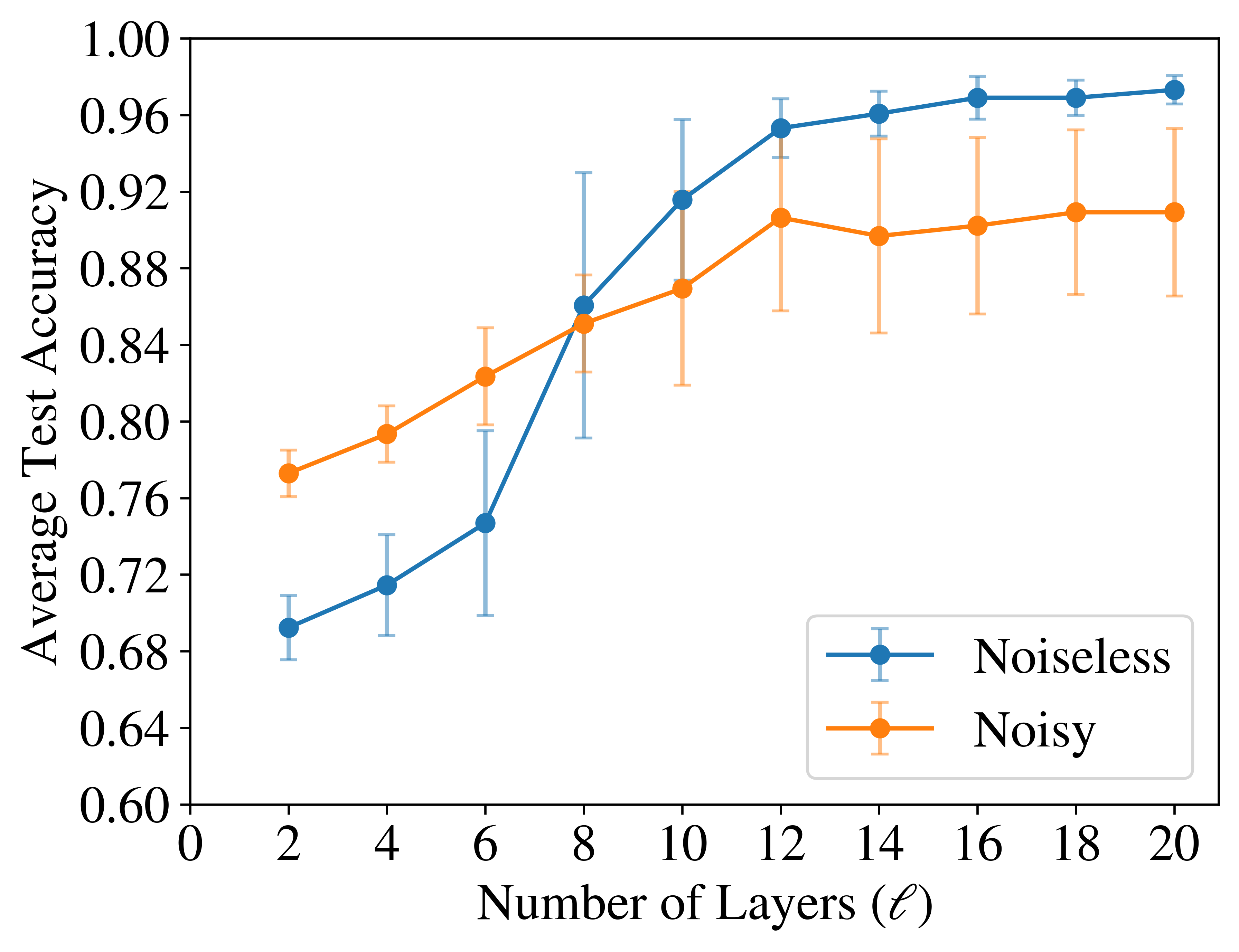}
        \caption{\figupdated{Comparison of the average test accuracy when using a digital circuit for the digit comparison of 1 versus 9.}}
        \label{fig:cnot1v9}
    \end{figure}

    Additionally, we see in Fig.~\ref{fig:cnot1v9} that in the noiseless 1 versus 9 comparison case, we do not see much improvement in the test accuracy after 16 layers, and coincidentally see that this is also optimal for the noisy model. To note the general case across digit comparisons, in Fig.~\ref{fig:compare-depth}, we see accuracy plateauing around $\ell = 16$ as well. However, $\ell = 16$ does not seem to be optimal for the noisy case in general as seen in Fig.~\ref{fig:compare-depth}.
    
    \subsubsection{Computational details}
    All instances of learning used the \texttt{AdaGrad} optimizer from \texttt{Flux} with a learning rate of 0.1 for the classical optimzation. During each training epoch, a randomized batch was used for training and to calculate the gradient. There are 70 total training epochs where one randomized sample was taken each time. Then for average accuracies taken in Fig.~[\ref{fig:binary_classification}], 50 trials were taken over random starting parameters. 

    Also, when making any measurements during the training or test phase, a new noisy circuit is generated such that the noise is independent for any measurements. No data from the test set is used to inform the training process during gradient descent.

    The details of numerical convergence in training are deferred to Fig.~\ref{fig:convergence_comp}(a) the final section of the Appendix.

\subsection{\label{app:subapp:phase}Phase detection}

\subsubsection{Phase diagram discussion}
\begin{figure}
    \centering
    \includegraphics[scale=0.4]{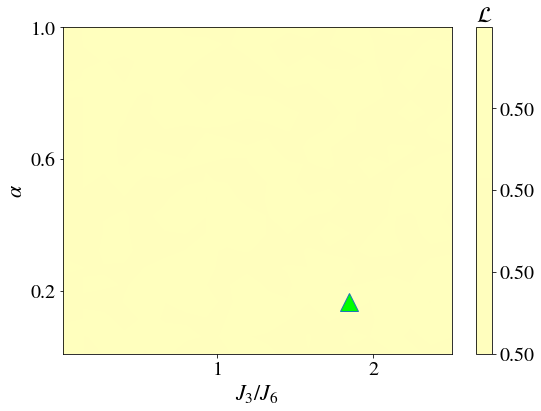}
    \caption{``Learned" phase diagram of the XXZ Hamiltonian using a $\ell = 0$ circuit (1-qubit gates only). The green triangle marks the training point.}
    \label{fig:xyz_rots}
\end{figure}

Two supplementary plots offer additional context for the phase learning previously described. First, the phase diagram of the XXZ Hamiltonian cannot be learned using a circuit with no entangling resource.  Figure \ref{fig:xyz_rots} depicts the result when such learning is attempted: uniform loss of 1/2 throughout the entire phase diagram.  This is in contrast to the Rydberg phase diagram, which could be learned using such a circuit, as seen in Fig.~\ref{fig:ryd_true_learned}(c).

\begin{figure}
    \centering
    \includegraphics[width=0.4\textwidth]{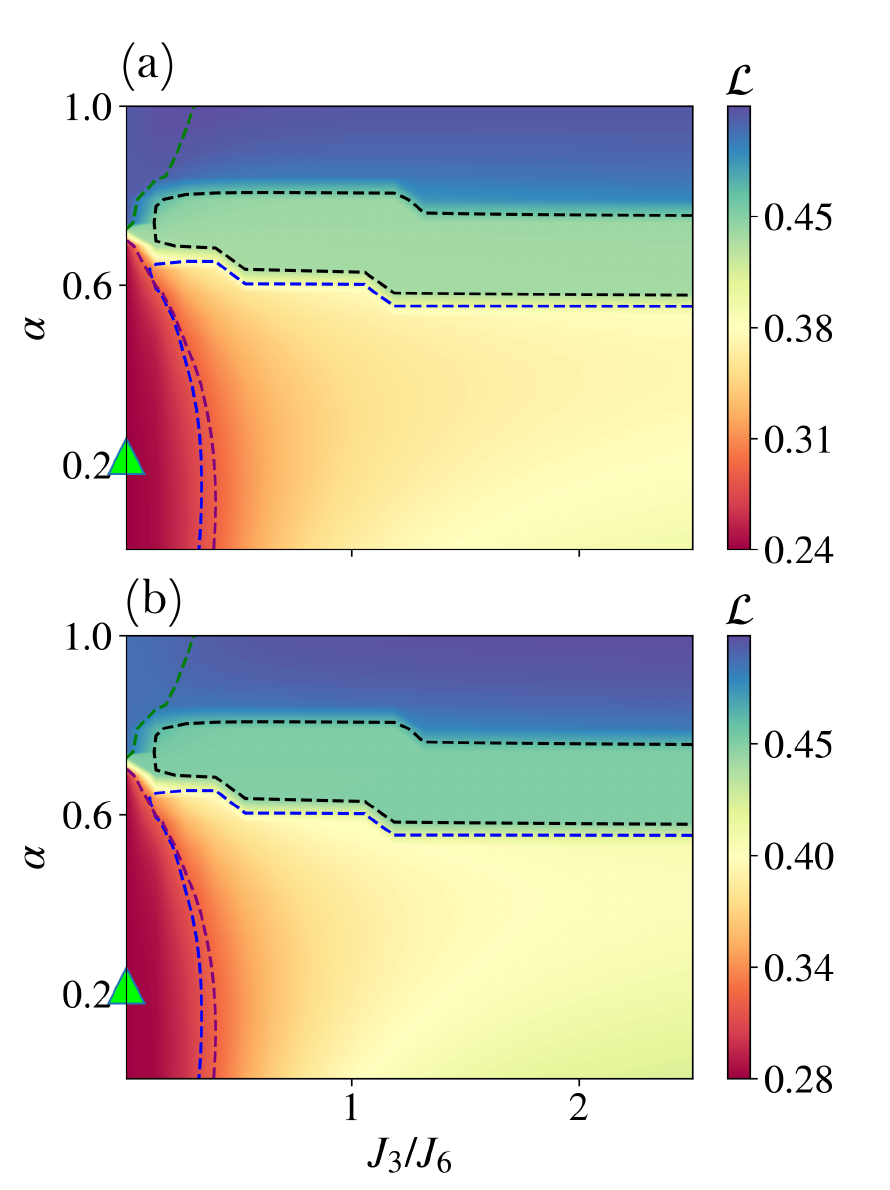}
    \caption{\figupdated{Learned phase diagrams for (a) digital-analog and (b) digital \updated{(with $\phi = \pi/4$)} circuits, with $n = 8, \ell = 2$ in the \updated{absence of noise}. Dotted contours overlay the true phase boundaries.}}
    \label{fig:depth2_noisycomp}
\end{figure}

In addition, in the \updated{noiseless} scheme we have asserted that learnability is not an interesting problem for comparison of digital versus DA circuitry.  This is because even at low depths, both circuits are able to carve out all four phases of the XXZ Hamiltonian: Figure \ref{fig:depth2_noisycomp}(a) demonstrates the DA-learned phase diagram with circuit depth 2, while Fig.~\ref{fig:depth2_noisycomp}(b) shows the digitally-learned diagram at the same depth. 
\begin{comment}
    before this was updated to be the correct noiseless d=2 plots, it was labeled_noisy_sidebyside_d2.png
\end{comment}

\subsubsection{Computational details}
Training was performed using a looped gradient descent technique, which at each step calculated the gradient of the loss function with respect to the circuit parameters.  This \updated{was} inputted into \texttt{Julia}'s \texttt{Flux Optimizer }~\cite{Flux.jl-2018, innes:2018}, which updated the circuit parameters at the end of each iteration.  Training in the Rydberg phases used 50 training iterations, whereas training in the XXZ Hamiltonian phases used 70 iterations for both noisy digital and noisy DA circuits\updated{, and 50 iterations for noiseless digital and DA circuits}.  For the Rydberg Hamiltonian, Fig.~\ref{fig:ryd_true_learned}(b) was trained at the $\mathbb{Z}_2$ phase point (2.5, 1.3538) and Fig.~\ref{fig:ryd_true_learned}(c) was trained at the disordered phase point (0.6, 1.3).  For the XXZ Hamiltonian, Fig.~\ref{fig:XXZ_LPD_4x4}(a) was trained on the zAFM phase point (0.01, 0.2184), Fig.~\ref{fig:XXZ_LPD_4x4}(b) on the qzAFM phase point 0.01, 0.9479), Fig.~\ref{fig:XXZ_LPD_4x4}(c) the XY-QLRO phase point (1.8447, 0.1663), and Fig.~\ref{fig:XXZ_LPD_4x4}(d) the VBS phase point (1.5826, 0.6353). The details of numerical convergence are shown in Fig.~\ref{fig:convergence_comp}(b) in the next section.

Testing was performed using an $n_x \times n_y$ discretized parameter mesh, where $n_x = n_y = 45$ for the Rydberg testing, and $n_x = n_y = 20$ for the XXZ Hamiltonian testing.  Training points were chosen directly from these meshes.  The phase diagrams presented in this work represent an average of the losses achieved over 20 separate training and testing runs.

One loss evaluation was performed in each iteration of the training loop (for DA, this was 70 iterations $\times$ 1 loss evaluation = 70 loss evaluations) and once per testing iteration (20 $\times$ 20 = 400 additional loss evaluations) for a total of 470 loss evaluations per run.  To achieve 1\% accuracy in the loss function that appears to be sufficient for reaching the performance here reported, a total of 10000 $\times$ 470 $\sim 5 \times 10^6$ samples were necessary.

\section{Numerical convergence}
\begin{figure}
    \centering
    \includegraphics[width=0.44\textwidth]{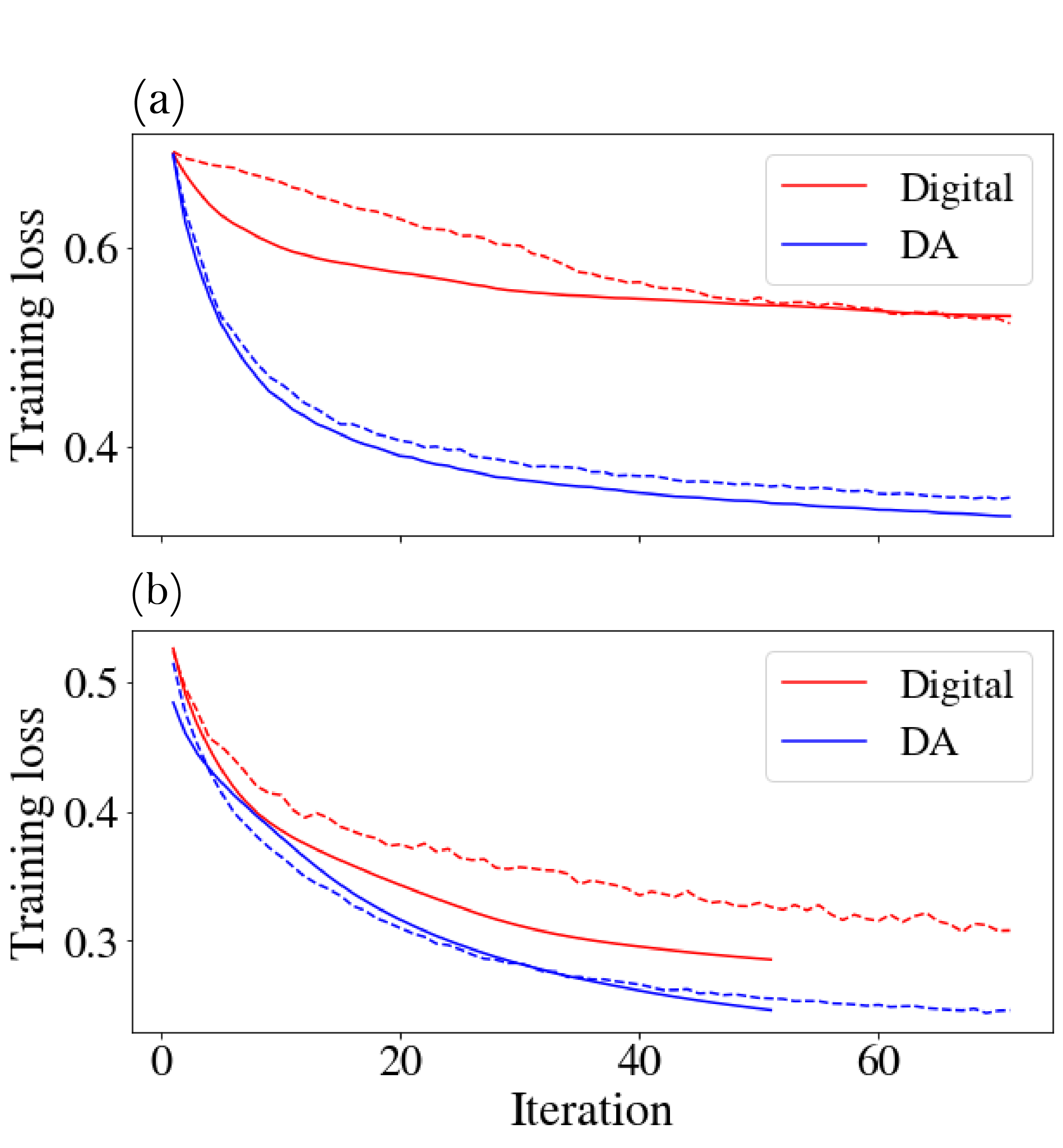}
    \caption{(a) 3 versus 8 digit classification training convergence for digital-analog (blue) versus digital (red) circuits in the noiseless (solid line) and noisy (dashed line) regimes. The number of qubits is $n = 9$, the number of layers is $\ell = 12$ and $R_b/a = 0.87$. (b) XXZ Hamiltonian training convergence in the zAFM phase for digital-analog (blue) and digital (red) circuits with $n = 8, \ell = 2$ in the noiseless (solid line) and noisy (dashed line) regimes. Each step represents the average loss over 20 runs.}
    \label{fig:convergence_comp}
\end{figure}

Figure~\ref{fig:convergence_comp} compares the training loss function at each iteration between digital \updated{(with $\phi = \pi/4$)} and digital-analog (red/blue), noiseless and noisy (dashed/solid), and on digit classification and phase learning (a/b). We obtained each curve by averaging over 20 random runs. In addition to the other potential improvements digital-analog learning has shown in the main text, Fig.~\ref{fig:convergence_comp} also demonstrates that digital-analog circuits may plausibly require fewer iterations to reach a desired threshold, though a more in-depth analysis would be necessary to more thoroughly investigate this possibility.

\end{document}